\documentclass[12pt,preprint]{aastex}
\usepackage{longtable}
\usepackage{pdflscape}

\slugcomment{Submitted to {\it The Astrophysical Journal}}

\shorttitle{The Afterglows of Short- and Long-Duration GRBs}
\shortauthors{Nysewander, Fruchter & Pe'er}

\def \etal{{\it et al.~}}

\newcommand{\cm}{\rm{\, cm}}
\newcommand{\dL}{\rm{\, d_{L,28}^{-2}}}
\newcommand{\mjy}{\rm{\, mJy}}
\newcommand{\Hz}{\rm{\, Hz }}

\newcommand{\beq}{\begin{equation}}
\newcommand{\eeq}{\end{equation}}
\newcommand{\ba}{\begin{array}}
\newcommand{\ea}{\end{array}}
\newcommand{\E}{E_{52}}

\newcommand{\n}{n_{0}}
\newcommand{\ee}{\epsilon_{e,-1}}
\newcommand{\eB}{\epsilon_{B,-2}}
\newcommand{\td}{t_{d}}
\newcommand{\Z}{(1+z)}
\newcommand{\Egi}{{\rm E}_{\gamma, \rm ISO}}

\begin{document}

\title{A Comparison of the Afterglows of Short- and Long-Duration Gamma-Ray Bursts}

\author{M. Nysewander\altaffilmark{1, 2}, A.S. Fruchter\altaffilmark{1} \& A. Pe'er\altaffilmark{1,3}}

\altaffiltext{1}{Space Telescope Science Institute, 3700 San Martin Dr, Baltimore, MD, 21218}
\altaffiltext{2}{Alion Science \& Technology, 1000 Park Forty Plaza, Durham, NC 27713; mnysewander@alionscience.com}
\altaffiltext{3}{Giacconi Fellow}

%% ABSTRACT

\begin{abstract}

  We present a comparative study of the observed properties of the
  optical and X-ray afterglows of short- and long-duration
  $\gamma$-ray bursts (GRBs).  Using a large sample of 37 short
  and 421 long GRBs, we find a strong correlation between the afterglow
  brightness  measured after 11 hours and the observed fluence of the 
  prompt emission.   Both the  optical (R band) and X-ray flux densites ($F_R$ and $F_X$) scale with the  $\gamma$-ray
  fluence, $F_{\gamma}$.    For bursts with a known redshift,
  a tight correlation exists between the afterglow flux densities at 11 hours (rest-frame) and the
  total isotropic $\gamma$-ray energy, $\Egi$: $F_{R,X} \propto {\Egi}^{\alpha}$, with
  $\alpha \simeq 1$.
  The constant of proportionality is nearly identical for long and short bursts, when $\Egi$ is obtained from the Swift data.
  %Therefore, for a given fluence, the afterglows of short GRBs are not
  %significantly dimmer than those of long GRBs in the optical and the
  %X-ray bands. 
  Additionally, we find that for short busts with $F_{\gamma} \ga10^{-7}$ erg cm$^{-2}$, optical afterglows
  are nearly always detected by reasonably deep early observations.   Finally, we show that the ratio $F_R / F_X$ has very similar 
  values for short and long bursts.
  These results are difficult to explain in the framework
  of the standard scenario, since they require that either (1) the
   number density of the surrounding medium of short bursts is
  typically comparable to, or even larger than the number density of long
  bursts;  (2) short bursts explode into a density profile, $n(r)
  \propto r^{-2}$ or (3) the prompt $\gamma$-ray
  fluence depends on the density of the external medium.
  We therefore find it likely that either basic
  assumptions on the properties of the circumburst environment of
  short GRBs or else the standard models of GRB emission must be
  re-examined.    We believe that the most likely solution is that the 
  ambient density surrounding typical short bursts is higher than
  has generally been expected:  a typical value of $\sim 1$ per cm$^{-3}$ is indicated.
  We discuss recent modifications to the standard
  binary merger model for short bursts which may be able to explain the 
  implied density.  
\end{abstract}

%% more keywords


%% INTRODUCTION

\section{Introduction}

Although the past decade has seen great progress in characterizing the
afterglows of long-duration GRBs (LGRBs), the afterglows of
short-duration GRBs (SGRBs) remain elusive.  The average fluence of an
SGRB is an order of magnitude fainter than that of an LGRB, and thus
while SGRBs have been localized by {\it Swift}, about ten long bursts
are localized for every short burst.  Optical afterglows of SGRBs have
been detected, but are notoriously dim, and large-aperture telescopes
are generally required to detect them.  Nearly half of all SGRBs have
a fluence less than 10$^{-7}$ ergs cm$^{-2}$ and none of these have a
detected optical afterglow.  SGRB X-ray afterglows are commonly
detected, but are faint and less likely to be rapidly localized with
their positions distributed in real-time, further complicating optical
ground-based observations.  As a result, only a few well-sampled,
robust optical SGRB light curves exist that span a wide range of time
or frequency.

The faintness of the afterglows of short GRBs (compared to the long
GRBs) has been attributed to properties of the progenitor and
circumburst environment.  While the LGRB-SN connection is strong, the
progenitors of SGRBs are less certain.  Observations have shown that
long-duration GRBs are associated with the deaths of massive stars
(\citealt{smg+03, hsm+03}; see \citealt{bw06} for a comprehensive
review).  However, a number of models exist for the progenitors of
SGRBs.  The leading model is a binary pair of compact objects, dual
neutron stars, or a neutron star and black hole, which merge in a
dramatic explosion causing a burst of $\gamma$-rays \citep{elp+89,
  npp92, lrg05} powered by accretion onto the newly formed compact
object.
 
It is commonly believed that the difference in the long and short progenitors
implies that they reside in different environments (for a recent review, see
\citealt{n07}).  Because LGRBs are associated with massive stars, they will
lie in or near dense regions of active star-formation.  SGRBs progenitors can
have long lifetimes and do not need to be in galaxies with active star-formation,
nor be associated with dense regions.  In fact, the observed diverse host
galaxy morphology is roughly consistent with the expected wide range of
progenitor lifetimes \citep{bpb+06}.

Two of the first detected SGRB afterglows, those of GRB 050509B
\citep{gso+05} and GRB 050724 \citep{bpc+05}, were found near bright,
giant elliptical galaxies.  GRB 050709 \citep{hwf+05} and GRB 051221A
\citep{sbk+06} both occurred in late-type galaxies with ongoing
star-formation.  However, within these late-type galaxies, SGRBs do
not necessarily trace areas of active star-formation as is the case
with LGRBs \citep{bkd02, fls+06}, rather, they can occur in low
luminosity outer regions of the galaxy (e.g. GRB 050709;
\citealt{ffp+05}, GRB 060121; \citealt{ltf+06}).  Some theories of the
formation of compact object binaries include a violent
supernova-induced {\it kick} that can offset them from their host
galaxy \citep{bsr+06} and may place them in these low density
environments. 

The low environmental densities into which SGRBs occur have an
observational implication.  Within the context of the standard
afterglow model \citep[e.g.][]{MR97, spn98} for observing frequencies
below the cooling frequency (and above the peak frequency), the
afterglow flux is proportional to the density of the external medium
to the power $1/2$.  Well-sampled, multiwavelength observations of
LGRB afterglows have suggested that at half a day, the cooling
frequency generally lies between the optical and X-ray frequencies
\citep[e.g.][]{gwb+98,hbf+99,pk02,sbk+06}.  It is therefore expected
that the differences in the density of the circumburst medium between
the long and short bursts would affect the late time optical emission,
which is below the cooling break.  (Note that according to the
standard model, the flux above the cooling break is independent of the
external medium density, $n$. Thus, the flux at the X-ray frequency
may not be affected by the difference in densities. See discussion in
\S\ref{sec:analysis} below). 

In addition to depending on the environment, we also expect the
absolute afterglow brightnesses to be proportional to the total energy
released in the burst.  The intrinsic energy release of short burst
progenitors is not well-known because the current redshift
distribution of SGRBs is fragmentary.  The average redshift of the
SGRB distribution is $<z>$ = 0.5 with E$_{ISO} \sim 10^{48} -
10^{51}$, compared to $<z>$ = 1.9 and   E$_{ISO} \sim 10^{49} - 10^{53.5}$
for LGRBs.  However, it is unclear to what extend these
differences are a result of observational biases and/or true intrinsic
differences in luminosity or collimation in comparison with LGRBs.  No
SGRB absorption redshifts have been measured, which is likely due to
the lower afterglow brightness and possibly a less dense external
medium on which to imprint strong absorption lines.  Therefore,
because all SGRB redshifts have been measured from optical emission
lines in host galaxies, the current sample of redshifts is limited to
$z \la 1$.  \citet{bfp+07} have noted that the lack of bright galaxies
in the error circles of a number of XRT localizations may suggest that
a number of short bursts without optical identifications may lie at $z
> 1$.  Finally we note that the E$_{ISO}$ values presented here are
either based upon the $\gamma$-ray fluence observed by the Swift Burst Alert Telescope (BAT)
or for the smaller non-Swift sample, scaled to the BAT sensitivity. Because the
BAT is only sensitive up to 150 keV, these values underestimate the
total prompt emission of all bursts.   However, this is most severe
on short bursts, which tend to have a harder spectrum than long
bursts.  We return to this point in more detail later.

In this paper we compare the optical and X-ray afterglows measured
after 11 hours (a standard {\it Swift} timescale) with the prompt
energy emission of both long- and short-duration $\gamma$-ray bursts,
and compare these results with afterglow theory.  We test the
predictions of the standard model with a large data set that spans
over six decades in energy. We find that SGRBs afterglows are very
similar to low-luminosity LGRBs afterglows.  Surprisingly, we find
that the afterglow brightnesses depend little upon the classification
of long or short; rather, the bursts scale primarily with the
high-energy prompt emission.  Our result is difficult to reconcile
with standard afterglow theory, and contradicts previous claims that
SGRB optical afterglows are dim due to the density of their
surrounding medium \citep{pkn01, n07, b07d}.

This paper is organized as follows: In \S \ref{sec:data} we detail the
steps taken to gather the data presented herein.  We compare the
prompt and afterglow emission of the two populations in \S 3.  In \S
4, we explore the assumptions of the standard model, under what
conditions the results are valid, and implications related to host
galaxies and intrinsic SGRB energies.  We show that the afterglow
  data implies that either (1) the environment of short GRBs is
  comparable to that of long GRBs, (2) SGRBs explode into a wind
  profile, or (3) the prompt emission fluence depends on the
  environment. In \S\ref{sec:statistics} we discuss the statistical
  significance of our data sample.  We summarize and draw final
  conclusions in \S \ref{sec:summary}. 

%% DATA

\section{Data}
\label{sec:data}

In this study, we include GRBs detected by all satellites, beginning
with GRB 970111, the first localization with follow-up observations
capable of detecting an afterglow, and ending over a decade later,
with GRB 071227.  To be included in the analysis, a GRB must have
published values of the prompt emission and had optical or X-ray
afterglow follow-up observations within a few days of the burst.  421
LGRBs and 37 SGRBs have been included in the sample, out of which 408
LGRB and 37 SGRB have optical and 299 LGRBs and 27 SGRB have X-ray
follow-up observations.  For each GRB, Tables~\ref{short_table} and
\ref{long_table} present the short and long prompt gamma-ray fluence,
X-ray afterglow flux at eleven hours, and optical $R$-band magnitude
at eleven hours (times here are measured in the observer's frame).  We
also include basic properties of the burst: the satellite and
frequency range of the prompt data and the redshift, if known.  The
data come from over 700 unique sources; these sources are listed in
the tables.  We thoroughly searched the literature to provide the most
accurate values.  Only the first page of Table \ref{long_table} is
presented in the printed version of this paper.  The entire table can
be found in the Electronic Supplement.

The decision of whether a burst is classified as {\it short} or {\it
  long} is based upon many factors in the prompt emission and not
merely on the burst duration.  When a burst's sub-type is
questionable, we adopt the classification published by the instrument
team.  Formal measures of T$_{90}$ do not always accurately
distinguish long from short, such as in cases where we observe a short
hard spike followed by a long, soft tail of emission (e.g. GRB 050724
or GRB 051227; \citealt{bcb+05,bgn+05}).  Some bursts, notably GRB
040924 and GRB 050925, have short durations but soft emission, hence
are likely the tail-end of the long-duration population \citep{huf+05,
  hbb+05}.  GRB 050911 has a complex BAT $\gamma$-ray light curve that
may be interpreted as short, but because it is not conclusively in
either category \citep{pkl+06}, it is not included. 

Our primary sources for GRB fluence and duration are published
catalogues, such as those for {\it Swift} \citep{sbb+08} or {\it HETE}
\citep{slk+05}, however some are taken from refereed papers or, if no
other source is available, from GCN Circulars.  If no error is given
in the original publication, we assume that the GRB is detected at the
3$\sigma$ level.  Because this data set includes bursts discovered by
many satellites, we list in tables 1 and 2 the frequency range in
which the prompt emission is observed.

All fluences have been transformed into the {\it Swift} 15--150 keV
fluence range by determining empirical transformations derived from
bursts that were detected by multiple satellites.  For all bursts with
detections by more than one satellite, we calculate the fluence ratio
between the two detectors, and the deviation about this ratio.  Due to
sparse data and lack of overlap in mission lifetimes, in many
instances it is necessary to perform multiple transformations to
arrive at the final {\it Swift} BAT fluence (e.g. using the {\it
  Ulysses} to {\it Swift} ratio in order to relate {\it INTEGRAL} and
{\it Swift}).  However, the additional scatter introduced in doing so
is included in the final ratio error.  We jointly calculate the
transformation for the similar {\it HETE} 25--100 keV, {\it Ulysses}
25--100 keV, and {\it BATSE} 20-100 keV bands, and find the ratio to
{\it Swift} to be 1.65$\pm$0.31.  The high-energy {\it Konus-Wind}
(typically 20--2000 or 20--10000 keV) and {\it RHESSI} (30--10000 keV)
fluences were similarly grouped and found to have a rough
transformation ratio of 0.37$\pm$0.23 into the {\it Swift} range.  Few
bursts (seven) had joint detections with {\it INTEGRAL} (20--200 keV);
these indicate a ratio of 0.95$\pm$0.68.  The {\it HETE} (30--400 keV)
ratio is calculated to be 0.63$\pm$0.32, and the {\it BeppoSAX} ratio
is 0.67$\pm$0.45.  The scaling and large additional errors introduced
by this method compensate for the additional scatter caused by
comparing bursts between different detectors.  If we restrict our
analysis to {\it Swift} bursts only, our results do not significantly
change. 

\begin{figure}
\plotone{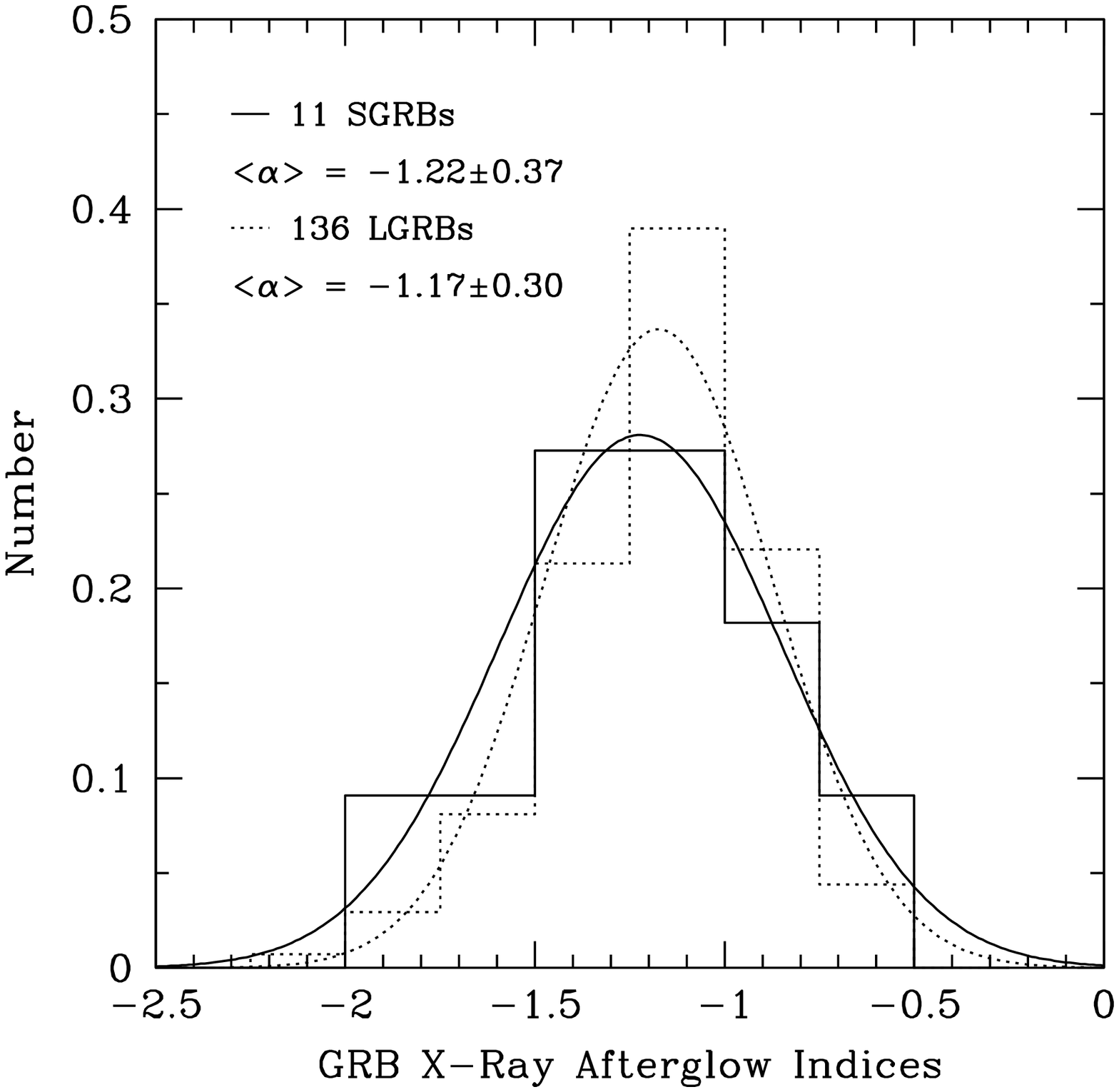}
\caption{A compilation of the X-ray afterglow temporal indices of the
  two populations and presents the best-fit Gaussian distribution to
  each data set.  We include only afterglow indices of bursts with a
  rapid XRT detection and with data extending to at least 11 hours.
  The solid line indicates SGRBs; the dashed indicates LGRBs.  The
  plot has been normalized by the sum of the bursts used: 11 short and
  136 long bursts. 
}
\label{xray_slope_histo}
\end{figure}

X-ray fluxes, corrected for Galactic extinction, are primarily taken
from large-scale uniform studies that fit light curves to estimate the
flux at eleven hours, such as those for {\it Swift} \citep{g+08} or
{\it BeppoSAX} \citep{dpg+06} GRBs.  For bursts not included in these
studies, we obtained data from refereed papers, the {\it Swift} XRT
Lightcurve Repository \citep{ebp+07}, or GCN Circulars.  Nearly all
X-ray afterglow fluxes are given in the 0.3--10 keV range, however a
few were reported in a different energy range and these are noted in
Tables 1 and 2.  For the few bursts with no reported Galactic extinction, we estimated the extinction from the neutral hydrogen column density using the Chandra Colden and PIMMS online calculators.  These bursts are noted in Tables 1 and 2.  In a few instances where no error is given, we
assume that the source is detected at the 3$\sigma$ level.  We choose
detections or limits as close to eleven hours as possible that we then
extrapolated using temporal slopes of $\alpha_{S, X}$ = $-1.22$ $\pm$
0.37 and $\alpha_{L, X}$ = $-1.17$ $\pm$ 0.30 for SGRBs and LGRBs
respectively.  These slopes are averages derived from power-law fits
that we performed of the light curves of all bursts observed by XRT
until December 31$^{st}$ 2007 that have a well-defined temporal slope
based on data that extend from minutes until at least eleven hours
after the burst.  Figure~\ref{xray_slope_histo} presents a histogram
of these slopes and corresponding Gaussian least-squares fit to the
scatter in the population.  The original measurement error and the
error associated with the extrapolation are added in quadrature to
determine the final error reported in the tables. 

\begin{figure}
\plotone{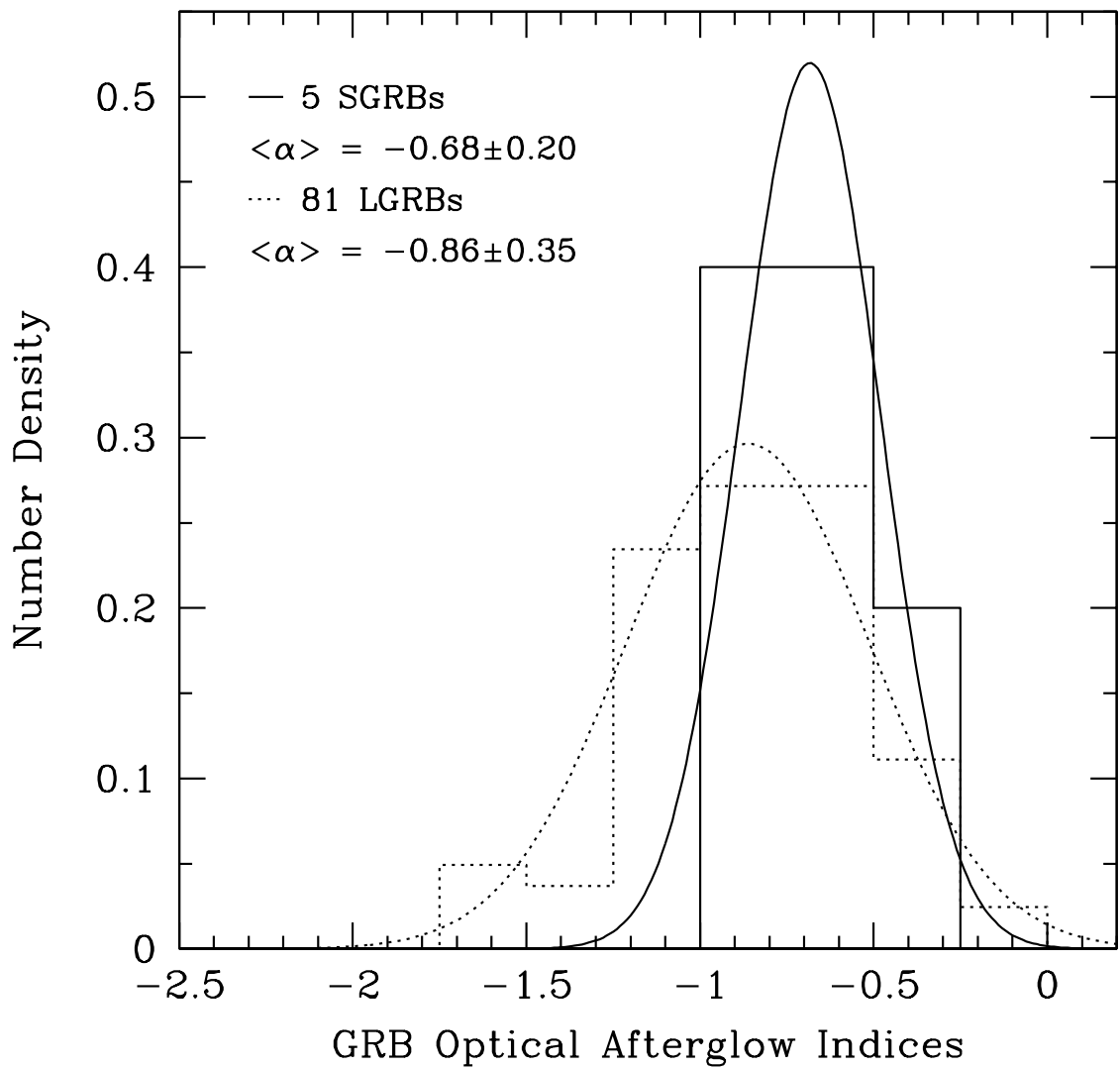}
\caption{A compilation of all short and long burst optical afterglow
  indices and the best Gaussian fits to the distributions.  We include
  only afterglow data that have light curves extending from at most
  three hours to at least eleven hours.  The solid line indicates
  SGRBs; the dashed indicates LGRBs.  The plot has been normalized by
  the sum of bursts used: 5 short and 81 long. 
}
\label{opt_slope_histo}
\end{figure}

Optical magnitudes are obtained from refereed papers when available
and from GCN Circulars when not.  We chose observations taken as close
to eleven hours as possible, preferably in the $R$-band.  All optical
magnitudes have been corrected for Galactic extinction using the
  method presented in 
\citet{sfd98}.  If no error is given, as is often the case for GCN
derived magnitudes, we again assume that the source is detected at the
3$\sigma$ level.  These data are extrapolated in time using a temporal
slope of $\alpha_{L, opt} = -0.85$ $\pm$ 0.35 for long and $\alpha_{S,
  opt}$ = $-0.68$ $\pm$ 0.20 for short GRBs, and in frequency using a
spectral slope of $\beta = -1.0$.  Because of a lack of early optical
data, the optical indices are averages based on all afterglow
observations with data that extend from at least three hours until
after eleven hours following the burst.  We fit power-laws to the
afterglows of the five SGRBs and 81 LGRBs that fit this selection
criteria.  Figure~\ref{opt_slope_histo} presents a histogram of these
slopes and Gaussian least-squares fit to the scatter.  The original
measurement error and the error associated with the extrapolation are
added in quadrature to determine the final error.

Often, multiple optical limiting magnitudes exist for a given GRB, and
in that case, we took care to choose the one that is most
constraining.  This sometimes can be arbitrary since the level of
constraint depends upon the intrinsic shape of the light curve and
redshift of the progenitor, both of which are unknowns given a
non-detection.  At early times, optical afterglows are often not
simple power-laws so  we avoided limiting magnitudes taken in the
first few minutes.  Additionally, we require that $>$95\% of the
satellite-localized error circle be covered by optical observations,
and we assume that the limiting magnitude of the DSS is $R$ = 20. 

The computed average optical indices are significantly shallower than
the X-ray indices.  One may expect an observational bias to be
introduced because we may be preferably selecting afterglows with
shallower temporal decays, because afterglows are typically more
difficult to observe in the optical than in the X-ray.
In order to show that our analysis is unbiased, we made the
  following test. We selected
%  However, if we select
 X-ray indices by the criteria of the GRB having an optical detection,
and through this, we applied the same optical selection effect
to the X-ray.  When making this selection, we found that the
average X-ray temporal slopes remain nearly unchanged ($-1.19$ versus
$-1.16$).  Therefore,  we conclude that the difference between
the optical and X-ray temporal indices is real and is not an
observational effect.

%% ANALYSIS

\section{Data Analysis}
\label{sec:data_analysis}

\begin{figure}
\plotone{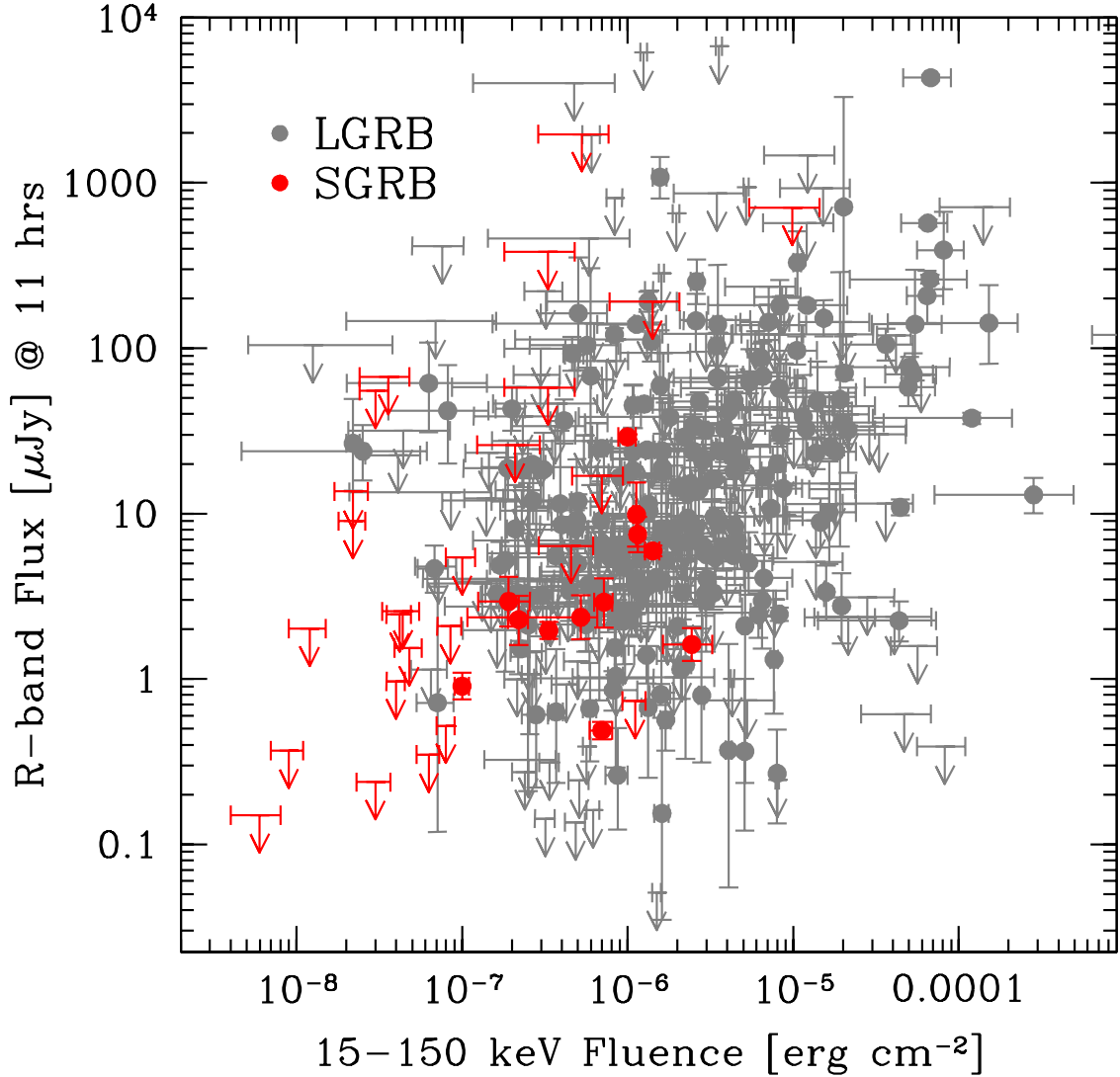}
\caption{A plot of the optical $R$-band flux (corrected for Galactic
  extinction) at eleven hours (observed frame) versus prompt 15--150 keV $\gamma$-ray
  fluence for both long (grey) and short (red) bursts.  Note that
  below a fluence of 10$^{-7}$ ergs cm$^{-2}$, no optical afterglow of
  an SGRB has been discovered, while above 10$^{-7}$, all reasonably
  deep observing campaigns but one (GRB 061210) have detected an
  optical afterglow.  }
\label{fig:opt_flux}
\end{figure}

\begin{figure}
\plotone{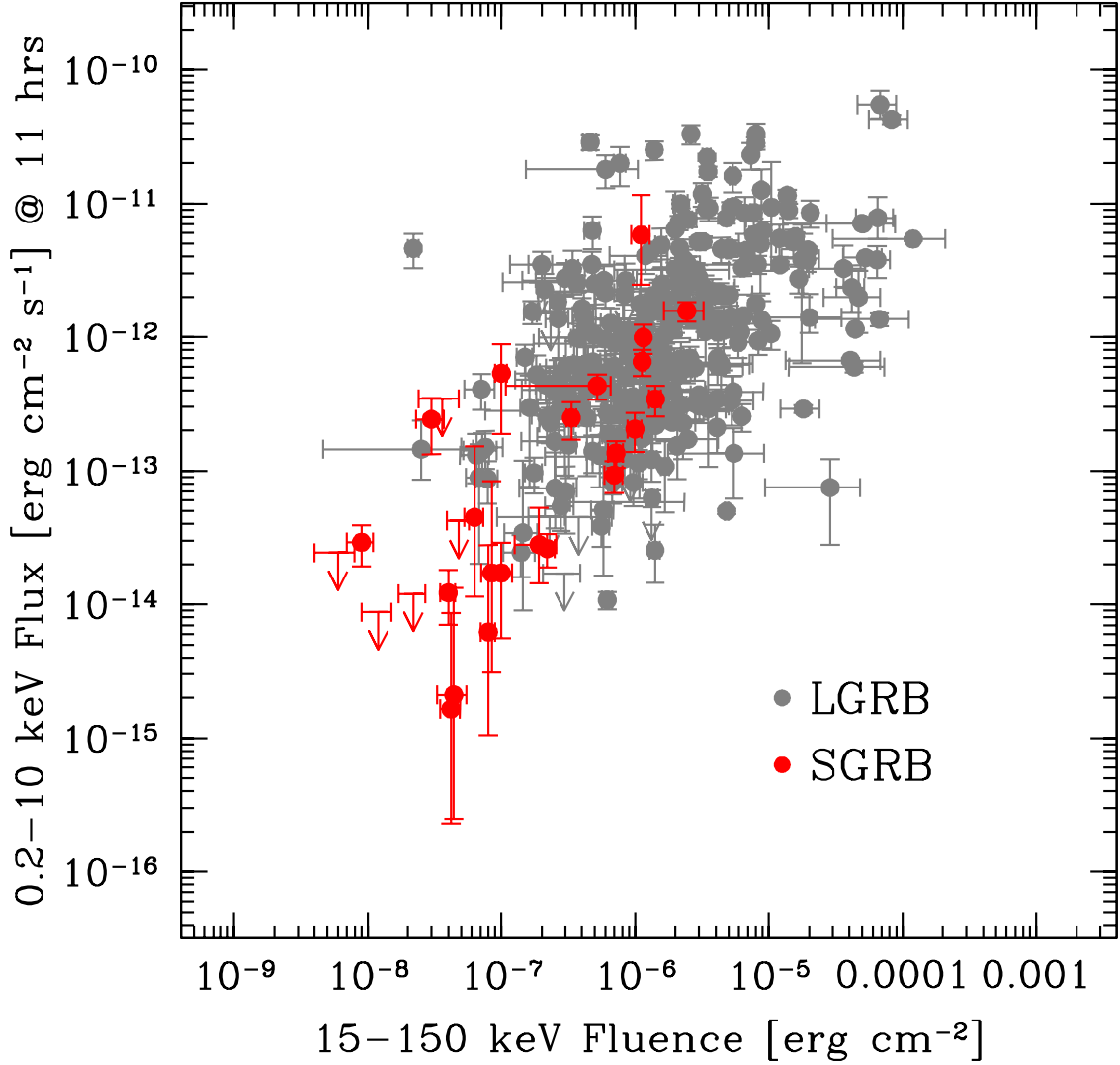}
\caption{A plot of the X-ray flux at eleven hours (observed
    frame) versus the prompt
  15--150 keV $\gamma$-ray fluence for both long (grey) and short
  (red) bursts.  }
\label{fig:X_flux}
\end{figure}

The entire sample of 421 long and 37 short bursts are presented in
Tables~\ref{short_table} and \ref{long_table} and are plotted in
Figures~\ref{fig:opt_flux} and \ref{fig:X_flux}.
Figure~\ref{fig:opt_flux} presents the optical brightness at eleven
hours (observer frame) versus the prompt high-energy fluence;
Figure~\ref{fig:X_flux} presents the X-ray brightness at eleven
hours (observer frame) versus the prompt fluence.  In both
figures, we see the same trend: GRBs with high fluence or energy have
greater brightness in both optical and X-ray.  Not unexpectedly, at
both high and low afterglow wavelengths, there is a large intrinsic
scatter in the data; the full width in the optical distribution spans
nearly four orders of magnitude.  A portion of the scatter is
certainly intrinsic to the population -- we do not expect tight linear
correlations.  However, the scatter in the optical is larger than in
the X-ray, and this may be partially due to observational effects.
The optical data comes from a wider variety of sources: different
telescopes, filters, instruments, observing teams, reduction and
calibration methods, etc. Line-of-sight source-frame absorption may
also play a role in the large optical scatter.

\begin{figure}
\plotone{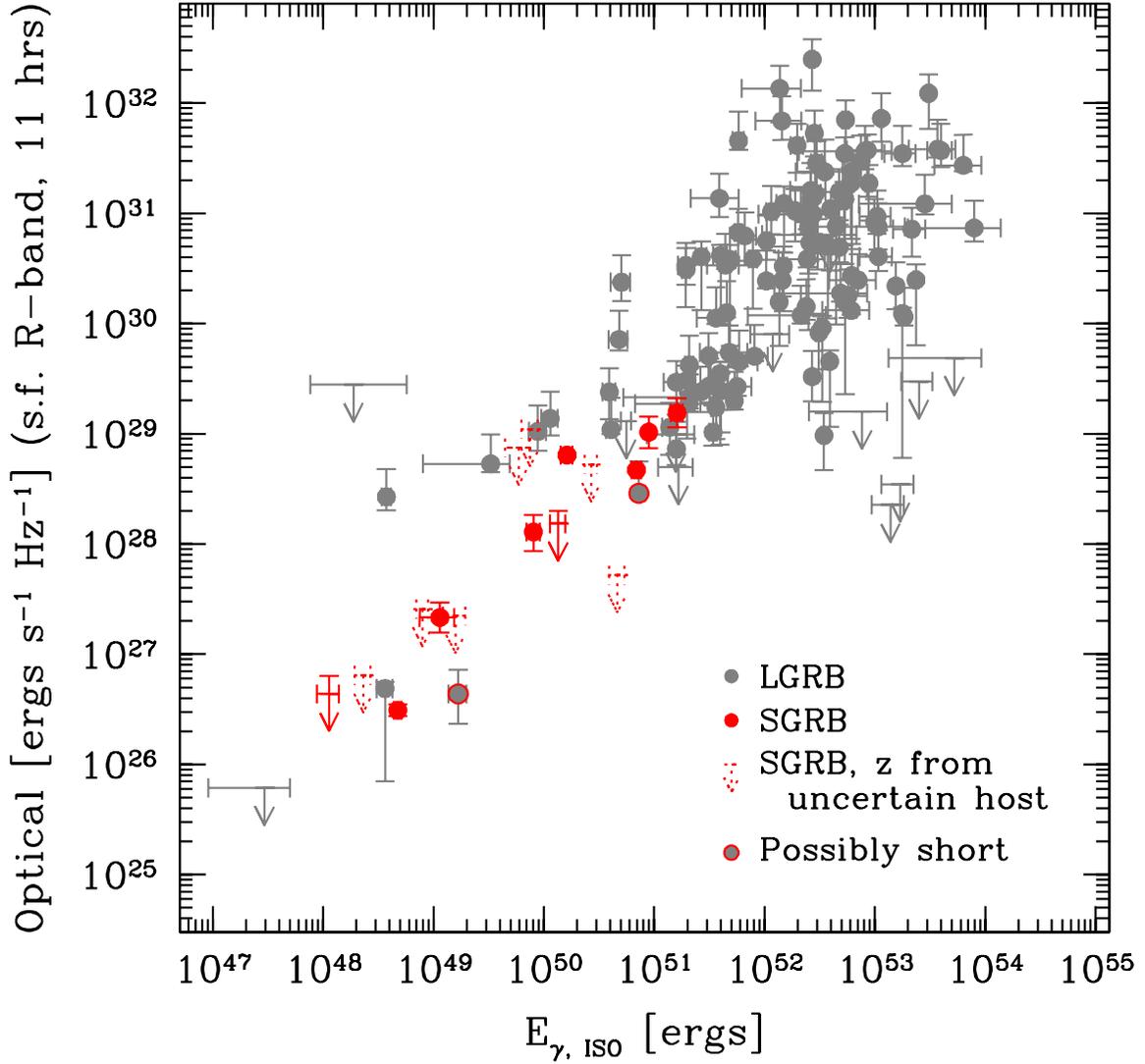}
\caption{A plot of the rest-frame optical $R$-band (corrected for
  Galactic extinction) afterglow brightness at eleven hours (in the
  source frame) versus $\Egi$, the total prompt emission of the burst
  in gamma-rays.  Dashed upper limits represent SGRBs with a host
  galaxy determined by XRT error circle only.  The classification of
  GRB 060614 and GRB 060505 is uncertain, therefore they are labeled
  as ``Possibly short''. }
\label{opt_energy}
\end{figure}

\begin{figure}
\plotone{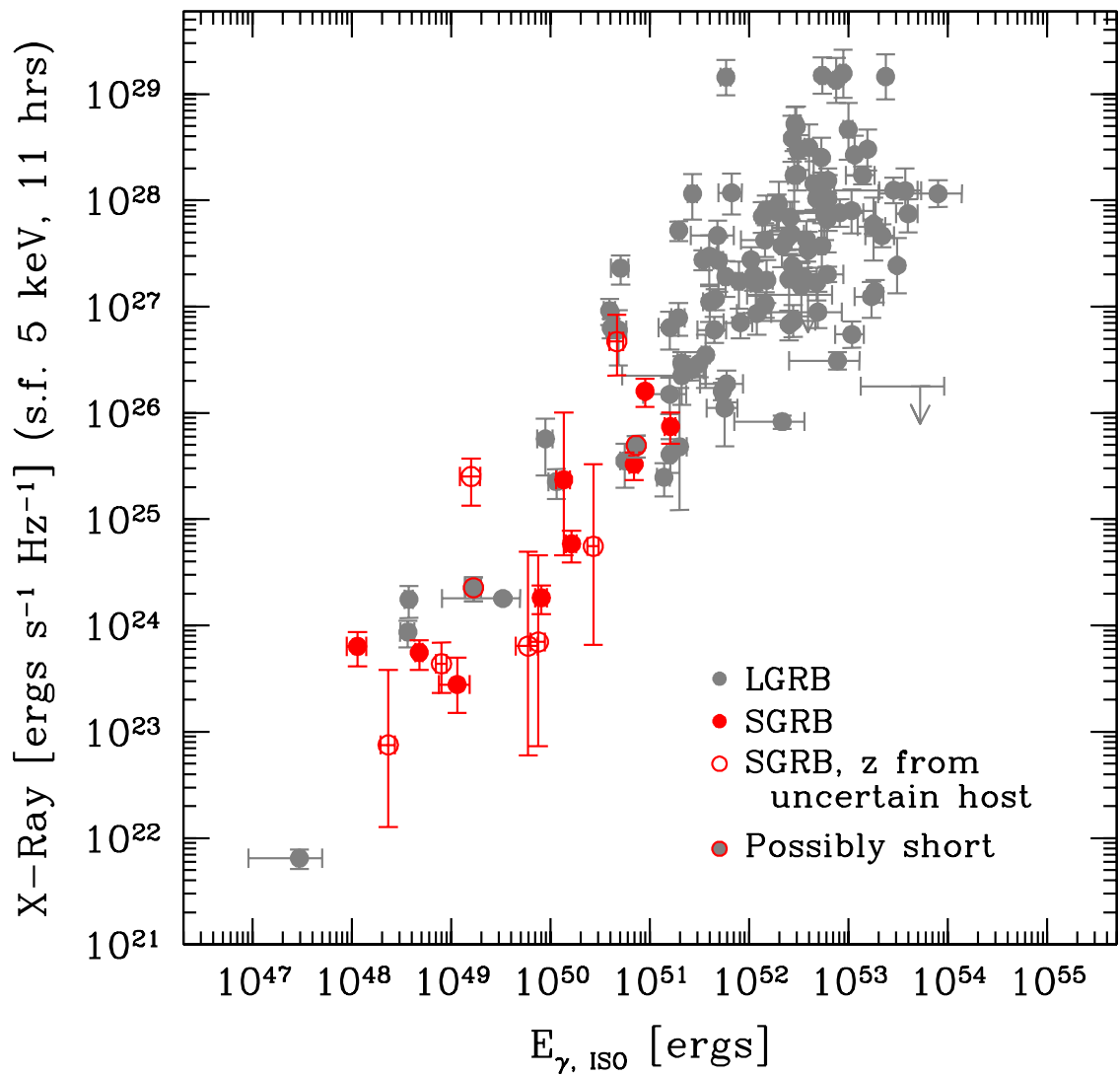}
\vspace{-0.7 in}
\caption{A plot of the rest-frame 5 keV X-ray afterglow brightness at
  eleven hours (rest frame) versus $\Egi$.   The open circles
  represent SGRBs with a host galaxy determined by XRT error circle
  only.  The classification of GRB 060614 and GRB 060505 is uncertain,
  therefore they are labeled as ``Possibly short.''  }
\label{xray_energy}
\end{figure}

The 126 long and 15 short bursts with measured redshift are presented
in the rest-frame in Figures~\ref{opt_energy} and \ref{xray_energy}.
Figure~\ref{opt_energy} plots the optical, while
Figure~\ref{xray_energy} plots the X-ray luminosities versus the
prompt   $\gamma$-ray energy, $E_{\gamma,{\rm ISO}}$.  As discussed
above, these plots exhibit a clear correlation between the afterglow
brightness at eleven hours and the prompt GRB fluence (note that in figures \ref{opt_energy},
  \ref{xray_energy} and \ref{fig:ratio} the time is measured in the
  source frame, and thus is redshift corrected with respect to the
  observed time).  Eight of the short
bursts have redshifts derived from hosts found with accurate optical
positions or are associated with Abell clusters, while seven have
hosts determined by XRT error circles only.  These seven bursts are
differentiated from the bursts with secure redshifts in
Figures~\ref{opt_energy} and \ref{xray_energy}, but are included in
the fit detailed below.  In calculating the isotropic equivalent
energy emitted in $\gamma$-rays, E$_{\gamma,{\rm ISO}} = 4 \pi
F_{\gamma} d_{L}^{2} (1+z)^{-1}$, the luminosity distance, $d_{L}$ is
determined assuming $H_{o} = 71$ km s$^{-1}$ Mpc$^{-1}$, $\Omega_{M} =
0.27$, $\Omega_{\Lambda} = 0.73$.  We transformed the optical data
from observer frame filters to rest-frame $R$-band using an assumed
spectral slope of $\beta = -1$.  The unabsorbed X-ray fluxes have been
transformed to 5 keV in the source frame assuming the integrated flux
has a spectral slope of $\beta = -1$ between 0.2--10 keV in the
observer frame.

While the existence of correlations for each of the two populations by
themselves is not surprising, it is clear from
Figures~\ref{opt_energy} and \ref{xray_energy} that the distributions
of the long and short bursts over-lap in both prompt and afterglow
energies.  In afterglow brightness, both the observed and rest-frame
populations overlap such that for a given fluence, the two classes are
not significantly different.  In each instance, both in optical and
X-ray, and in observer and rest-frame, the afterglow brightnesses
scale roughly as a power-law with energy.  Therefore, we model the
data as a one-dimensional Gaussian distribution about a straight line
in log-log space: $\log(F_{R,X}) = a + \alpha \log (E_{\gamma,{\rm
    ISO}})$.  We take into account errors in both coordinates by
minimizing the $\chi^2$ merit function:
\begin{equation}
\chi^2(a,\alpha) = \displaystyle\sum_{i=1}^n \frac{(y_i - a - \alpha
  x_i)^2}{\sigma^2_{y i} + \alpha^2 \sigma^2_{x i}}
\end{equation}
where we have fit to the line $y(x) = a + \alpha x$.  We note that
physically, $y(x) \equiv \log(F_{R,X})$, $x \equiv \log(E_{\gamma,{\rm
    ISO}})$, and $\sigma_{x i}$ and $\sigma_{y i}$ are the $x$ and $y$
standard deviations for the $i$th point.  The errors alone
over-constrain the linear fit and produce poor $\chi^2$ merit values.
Therefore, in order to compensate for the intrinsic width of the
distribution, we add a constant $1/3$ dex error to each error bar to
mimic the scatter effect.  Increasing the assumed population width
past $1/3$ dex does not significantly affect the final best fits.  Our
fits using this technique can be found in Table 3.  The slopes and
intercepts of the long and short burst distributions are equivalent,
within our ability to fit them.  While the small number of short
bursts means that the errors on a combined slope and intercept fit are
large, if we restrict the slope to $\alpha = 1$, then the intercepts
({\it i.e.} the brightnesses) of the two distributions are found to be
the same well within the error of $\sim 0.2$ dex.

The X-ray sample is dominated by observations performed by {\it
  Swift}, and therefore provides a largely homogeneous sample for our
analysis.  However, the interpretation of the optical data is hindered
by the vast diversity of follow-up programs and typical depths of
searches.  Many optical limits exist for bursts which do not constrain
the distribution, while some are deeply constraining.  Without knowing
the intrinsic source-frame extinction, it is not possible to determine
whether these reflect the intrinsic scatter in the distribution or if
they are a result of line-of-sight effects.  Limits of X-ray
afterglows are rare and often due to a delayed follow-up program
rather than intrinsic dimness of the afterglows.  Therefore, we do not
fit to limits in our analysis in either optical or X-ray.  However,
ignoring limits may introduce a selection bias to the fits. 

Figures~\ref{fig:opt_flux} and \ref{fig:X_flux} plot observer-frame
quantities and therefore have not been corrected for distance.  Given
a broadly homogeneous population, we expect the effect of distance to
dominate the slope of the populations.   In a Euclidean universe we
would expect the slope imposed to be $\alpha=1$.   The real universe,
of course, is expanding, and there is a
factor of $1+z$ difference between the correction of a bolometric
luminosity and a flux.   However, if, as is frequently found, the
spectral slope of the afterglow is $\beta \sim -1$, this factor of
$1+z$ is largely canceled by the k-correction.   And this may
largely be responsible for the excellent consistency between
the populations seen in the Figures and Table 3.

In Figures~\ref{opt_energy} and \ref{xray_energy} we show the subset of
bursts with published redshifts.  Compared to the previous plots, the
scatter in both figures has been reduced.  It is clear from both
figures, that the dominant predictor of afterglow brightness is the
prompt energy emission.  The fits to the optical and X-ray flux, given
in Table 4 are quite consistent between long and short bursts.  If we
again assume that the afterglow luminosities scale directly with energy, and
impose a slope of $\alpha=1$ on the rest-frame distributions, we find
that the long bursts are only brighter on average than the short
bursts for a given $E_{\gamma,{\rm ISO}}$ by about a factor of two,
while the long bursts themselves vary in brightness in the X-ray and
optical by orders of magnitude for a given $E_{\gamma,{\rm ISO}}$.
However, this difference is only at the $\sim 1 \sigma$ level and
thus may be entirely statistical rather than real.  We note here
that similar results to those discussed so far have been
derived by two other groups working contemporaneously, \citet{g+08} and \citet{Kann08}.

Readers may be concerned by the fact that we have not corrected the Swift data 
for the loss of total fluence in the $\gamma$-ray  due to the rather low
high-energy cutoff of the Swift BAT -- 150 keV.   This is well below the 
peak of the spectral energy distribution of most short bursts, although it is above 
the peak for most LGRBs with $E_{\gamma,{\rm ISO}} \la 10^{52.5}$~ergs \citep{amati06}.
Thus we may be underestimating the true $\gamma$-ray fluence of the 
short burst compared to the long bursts in some cases by up to a factor several \citep{Kann08},
particularly at low values of $E_{\gamma,{\rm ISO}}$  
where the two types of bursts overlap.
However, this correction would affect the plots of the afterglow flux density
in the optical and X-ray versus $E_{\gamma,{\rm ISO}}$  similarly.
As we will show in the next section, it is the ratio of the optical afterglow flux density to that in the X-ray
which provides this paper's real insight into the density of the media
surrounding the bursts, and this ratio is unaffected by corrections to $E_{\gamma} or E_{\gamma,{\rm ISO}}$ .

\section{A Comparison of Theory with Observations}
\label{sec:analysis}

\subsection{The Standard Model}
In order to compare the results presented in the previous section to
the predictions of the standard model of GRB afterglow emission, we
first give a short description of the model's predictions \citep[For a
much more detailed introduction to the theory see,
e.g.,][]{spn98}. Afterglow emission follows the propagation of a
relativistic shock wave that moves through the cold ambient medium,
whose density is assumed constant, $n=1 n_0 {\rm \, cm^{-3}}$. A
canonical value of $n_0 = 1$ is often used in the literature for
LGRBs, but values fitted from afterglow light curves and spectra of
long GRBs range from $\sim 10^{-3}$ to $\sim 10^{2}$
\citep[e.g.,][]{pk02,sbk+06}.  The shock wave moves in a self-similar
motion \citep{BM76}, which implies shocked fluid Lorentz factor
$\Gamma(E,n,R) = (17E/16\pi n m_pc^2 R^3)^{1/2}$. Here, $E\equiv
E_{\rm ISO}$ is the total explosion energy \footnote{Note the
  difference between the total explosion energy $E_{\rm ISO}$, and the
  energy emitted in $\gamma$-rays, $E_{\gamma,{\rm ISO}}$.}, and $R$ is
the radius of the shock wave, which is related to the observed time
via $t \approx R/4\Gamma^2 c$ \citep{W97}.

The shock wave accelerates electrons and generates magnetic
field. The accelerated electrons assume power
law distribution $n_{el}(\gamma) d\gamma \propto \gamma^{-p} d\gamma $
above $\gamma_m = \epsilon_e (m_p/m_e) g(p) \Gamma$. Here,
$\epsilon_e$ is the fraction of post-shock thermal energy that goes
into the electrons, $\Gamma$ is the Lorentz factor of the shocked
fluid, and $g(p)$ is a function of the power law index of the
accelerated electrons, $g(p) = (p-2)/(p-1)$ for $p>2$, $g(p) =
\log(\gamma_{\max}/\gamma_m)^{-1} \approx 1/7$ for $p=2$, where
$\gamma_{\max}$ is the maximum Lorentz factor in which power law
distribution of the electrons exist \citep[e.g.,][]{PW05}. A fraction
 $\epsilon_B$ of the post shock thermal energy is assumed to be
 converted to the magnetic field. With these definitions,
synchrotron flux peaks at observed frequency
\beq
\nu_m^{ob} = {3 \over 4 \pi} \Gamma \gamma_m^2 {q B \over m_e c} = 7.8
\times 10^{12} g(p)^2 \Z^{1/2} \; \E^{1/2} \, \ee^2 \, \eB^{1/2} \,
\td^{-3/2} \, \Hz,
\label{eq:nu_m}
\eeq
where $Q=10^x Q_x$ in cgs units was used, and the observed time is
measured in days.

The peak of the observed spectral flux at $\nu_m$ is
\beq
F_{\nu,\max}^{ob} = {1 \over 4\pi d_L^2} N_e P_{\nu_m} \approx {N_e \over
  4\pi d_L^2} {P_{tot} \over \nu_m^{ob}} = 7.3 \, \Z \dL \, \E \, \eB^{1/2} \,
\n^{1/2} \, \mjy,
\label{eq:F_nu_m}
\eeq
where $P_{tot} = \Gamma^2 (4/3) c \sigma_T \gamma_m^2 (B^2/8\pi)$ is
the total synchrotron power radiated by electrons at $\gamma_m$,
$\sigma_T$ is Thomson cross section, and $N_e=4 \pi R^3 n / 3$ is the
total number of radiating electrons. In estimating the observed flux,
luminosity distance $d_L = 10^{28} d_{L,28} \cm$ is assumed.

The synchrotron peak flux frequency is below the break frequency
$\nu_c$, the characteristic synchrotron emission frequency of
electrons for which the synchrotron cooling time, $6 \pi m_e c/B^2
\sigma_T \gamma_c (1+Y)$ is comparable to the ejecta (rest frame) expansion
time, $\sim r/\Gamma c$,
\beq
\nu_c^{ob} = 4.0 \times 10^{15}(1+Y)^{-2} \Z^{-1/2} \; \E^{-1/2} \, \eB^{-3/2} \,
\n^{-1} \, \td^{-1/2} \, \Hz.
\label{eq:nu_c}
\eeq
Here, $Y$ is the Compton parameter. As we show below, for the
canonical parameters values used here, after 11 hours $Y \ll 1$. Therefore,
the inclusion of $Y$ does not change the parametric dependence of $\nu_c$.

The results in equation \ref{eq:nu_m} indicate that the peak frequency
is below the optical band ($\nu_R = 5 \times 10^{14} \Hz$) after 11
hours, even for equipartition values of $\epsilon_e$ and
$\epsilon_B$. Since $\nu_c > \nu_m$, (the ``slow cooling'' regime),
the flux at frequencies above $\nu_m$ is given by
\beq
F_\nu^{ob} = \left\{
\ba{ll}
F_{\nu,\max}^{ob} (\nu/\nu_m)^{-(p-1)/2}, & \nu_m<\nu<\nu_c, \nonumber \\
F_{\nu,\max}^{ob} (\nu_c/\nu_m)^{-(p-1)/2} (\nu/\nu_c)^{-p/2}, & \nu_c < \nu.
\ea
\right.
\label{eq:F_nu}
\eeq

\subsection{The Comparison}
Two of the primary observational results obtained in this paper, summarized
in  figures \ref{opt_energy} and \ref{xray_energy},
can now be compared to the predictions in equations \ref{eq:nu_m} --
\ref{eq:F_nu}. For bursts with known redshift, $E_{\rm ISO}$ and
observed time $t_d = 11/24$, adopting the assumption that $2\leq p
\leq 2.5$, there are 3 free parameters whose values
are unknown: the ambient medium number density $n$ and the parameters
$\epsilon_e$ and $\epsilon_B$. In order to remove some of the
degeneracy, we look at the ratio of the fluxes at the $R$ and the $X$
bands. This ratio for the long and short GRBs is presented in
Figure~\ref{fig:ratio}.

While the peak flux frequency is below the optical (R) band, the value
of the cooling break depends on the density of the ambient medium and
on $\epsilon_B$.
Thus, three options exist:
\begin{enumerate}
\item The break frequency is above the observed X-ray frequency,
  $\nu_R < \nu_X = 1.25 \times 10^{18} \Hz < \nu_c^{ob}$.  In this
  case the ratio of the fluxes at the R and X-ray bands is predicted
  by equation \ref{eq:F_nu} to be $F_R/F_X = (\nu_x/\nu_R)^{(p-1)/2} =
  50$ (for p=2). This ratio is inconsistent with the results presented
  in Figure \ref{fig:ratio}, which clearly indicate a ratio $F_R/F_X
  \simeq 1-2 \times 10^3$ for both the short and long bursts. The flux
  ratio increases with $p$, and for $p=2.5$ it is $\sim 350$. Still,
  this is lower than that observed for the vast majority of
  bursts, and is more than an order of magnitude below the ratio of
  many.
 
\item The break frequency is in between the optical and the X-ray
  bands, $\nu_R < \nu_c^{ob} < \nu_X$.  As noted earlier many studies
  of afterglows from long GRBs have found this to be the case
  \citep[e.g.][]{gwb+98,hbf+99,pk02,sbk+06}. Indeed, in this
  situation, the expected ratio of the fluxes, $F_R/F_X =
  (\nu_x/\nu_R)^{(p-1)/2} (\nu_x / \nu_c)^{1/2} = 720 \, \E^{1/4}
  \eB^{3/4} \n^{1/2}$ (calculated for $\nu_c$ after 11 hours and for
  $p=2$) is consistent with the results obtained in Figure
  \ref{fig:ratio} for the long bursts. For $p=2.5$, this ratio is
  somewhat higher $\approx 5 \times 10^3 \, \E^{1/4} \eB^{3/4}
  \n^{1/2}$, which is slightly above the results in Figure
  \ref{fig:ratio}.

  On the other hand, for characteristic energy $E \simeq 10^{50}$~erg
  typical for the short bursts, this ratio is $F_R/F_x \simeq 130 \,
  E_{50}^{1/4} \eB^{3/4} \n^{1/2}$ (for $p=2$),  nearly an order of
  magnitude lower than the result obtained in Figure \ref{fig:ratio}
  for the short bursts. For $p=2.5$, the obtained value is $F_R/F_x
  \simeq 1600 \, E_{50}^{1/4} \eB^{3/4} \n^{1/2}$, consistent with the
  results in this figure. The value of $ F_R/F_X \approx 1600$ is
  obtained for values of the ambient number density of short bursts
  which are comparable to that expected for long bursts, $n_0 =
  1$. Lower values of the number density would be inconsistent with
  the results in Figure \ref{fig:ratio}, unless the value of
  $\epsilon_B$ is larger for short bursts than for the long ones.
  This discrepancy is further aggravated by observed $F_R/F_x$ values
  around $10^4$ for several short bursts.  It is hard to make these
  consistent with the theory using any set of parameters consistent
  with standard expectations.

\item The third possibility is that the cooling frequency is below the
R band, i.e., $\nu_c^{ob} < \nu_R < \nu_X$. Using equation
\ref{eq:F_nu} one obtains the ratio of the R to X-ray fluxes to be
$F_R/F_X = (\nu_x/\nu_R)^{p/2} = 2500$ (for $p=2$), which is
consistent with the results presented in Figure \ref{fig:ratio}. On the
other hand, the requirement that $\nu_c^{ob} <\nu_R$ for short GRBs for
which $E=10^{30}$~erg implies $\n > 130 \eB^{-3/2}$. Even for
equipartition value of the magnetic field this requirement implies
value of $n$ which is at least as high for short GRBs as for the
long GRBs.
\end{enumerate}

The analysis presented above implies that under the assumption of the
synchrotron emission model, the characteristic values of the number
densities of short GRBs are required to be at least similar to the
characteristic values obtained for long GRBs, $n \gtrsim 1 {\rm \,
  cm^{-3}}$. Lower values of the number density in short bursts
require stronger magnetic field. Moreover, the results hint toward
power law index $p \simeq 2.5$ in short bursts.

The ratios calculated above are done in the framework of the
  standard model, which assumes knowledge of total explosion energy
  $E_{\rm ISO}$, while for bursts with measured redshift we only
  measure the energy emitted in $\gamma$ rays, $E_{\gamma,{\rm
      ISO}}$. However, as we show in \S\ref{sec:energy_comparison}
  below, the efficiency in converting the explosion energy into
  $\gamma$ rays is very high, $E_{\gamma,{\rm ISO}} / \epsilon_e
  E_{\rm ISO} \simeq 1$ for both the long and short bursts.
  We note that in principle it is possible that the average values of
  the microphysical parameter $\epsilon_e$, $\epsilon_B$ and the power
  law index $p$ are different for short and long GRBs, which may
  affect our result. However, we find this possibility unlikely, since
  after 11 hours both types of bursts are well in the self-similar
  phase, in which the properties of the blast wave - that determines
  the values of these parameters - are well defined. 

The basic results found do not change if one assumes that the flux at
the X-ray band is dominated by Compton scattering \citep[see,
e.g.,][]{SE01, ZM01}. For $\nu_c^{ob} < \nu_X$ and $\nu_m^{ob}
\gamma_m^2 < \nu_X < \nu_c^{ob} \gamma_m^2$, one obtains the ratio of
the Compton to the synchrotron flux at the X-band to be
$F_{X,IC}/F_{X,syn} \approx (16/3) \sigma_T n r \gamma_m^{p-1}
(\nu_x/\nu_c)^{1/2} = 2.8 \times 10^{-2} g(p) \, \E^{5/8} \ee
\eB^{3/4} \n^{9/8} \td^{1/8}$, where $p=2$ is assumed. This ratio
clearly indicates that inverse Compton emission does not dominate the
flux at the X band unless the number density is significantly higher
than $n_0=1$.  Therefore, addition of IC component does not ease the
requirement for high values of the number density in short GRBs.

We have so far assumed that the density of the ambient medium is
constant, i.e. independent of radius from the burst. This, however, may not be the case.
Motivated by the association of long GRBs to the deaths of massive
stars, some authors \citep[c.f.][]{Woosley93, Pac98} have proposed
that the explosion producing a GRB may occur into a wind, ejected by
the progenitor prior to its explosion. This scenario results in an
ambient matter density profile, $\rho(r) = A r^{-2}$, with
characteristic value $A = 5 \times 10^{11} A_* {\rm \, gr \, cm^{-2}}$
typical for Wolf-Rayet star \citep{CL00}.

In an explosion into density profile, the evolution of the Lorentz
factor of the flow in its similar expansion phase is $\Gamma (E, r) =
(9 E / 16 \pi \rho c^2)^{1/2} r^{-3/2}$, and the relation between the
radius of the shock front and the observed time is $t \approx r / 2
\Gamma^2 c$ \citep{PWi05}. Similar calculations to the ones presented
in equations \ref{eq:nu_m} -- \ref{eq:nu_c} results in \citep[see,
e.g.,][]{PK01}
\beq
\ba{lcl}
\nu_m^{ob.} & = & 1.6 \times 10^{11}g(p)^2 \, \Z^{1/2} \E^{1/2} \ee^2
\eB^{1/2} \td^{-3/2} \, \Hz, \nonumber \\
\nu_c^{ob.} & = & 1.4 \times 10^{14} \,(1+Y)^{-2} \Z^{-3/2} \E^{1/2} A_*^{-2}
\eB^{-3/2} \td^{1/2} \,\Hz,  \nonumber \\
F_{\nu,\max}^{ob} & \simeq & 45 \Z \dL \E^{1/2} A_* \eB^{1/2}
\td^{-1/2} \, \mjy.
\ea
\eeq
Focusing on the case $\nu_m^{ob} < \nu_R < \nu_c^{ob} < \nu_X$, using similar
arguments to the ones presented for the constant density scenario
discussed above imply that the observed ratio $F_R/F_X \simeq 10^3$ is
obtained for combination of the parameters which fulfill $\E^{1/2}
A_*^{-2} \eB^{-3/2} \approx 33$. For long GRBs characterized by $E
\simeq 10^{52}$~erg and canonical value of $\epsilon_B = 0.01$, the
obtained value of $A_*$ is $A_* \approx 0.15$. For short GRBs for
which $E \approx 10^{49}$~erg, the obtained value is lower, $A_*
\lesssim 0.03$. The assumption $\nu_c^{ob} > \nu_X$ is
inconsistent with the results presented in Figure \ref{fig:ratio} and the
assumption $\nu_c^{ob} < \nu_R$ leads to higher values of $A_*$.

We thus find it possible to explain the results found for short GRBs
in a scenario in which the explosion that produces short GRBs occurs
into a wind profile, with wind density which is $\sim 30$ times lower
than the typical value for Wolf-Rayet stars. There are both strong
theoretical \citep[c.f.][]{CL00} and observational \citep{shr+08}
reasons to believe that a substantial fraction of long GRBs would
explode into a medium with a wind structure.  However, the standard
models of SGRB production involve the merger of compact binaries.  The
mass loss from the supernovae producing the compact remnants should,
however, impart velocities to the binaries that would take them far
outside of any remnant stellar winds long before they merge.
Therefore, if a wind density profile is indeed a key to understanding
the emission of SGRBs, a novel progenitor system will be required.

\subsection{A Further Limit on Emission Mechanisms}
\label{sec:energy_comparison}

The results presented here also place strong constraints
on the efficiency of $\gamma$-ray production during
prompt emission, significantly extending the limits
derived in earlier works \citep{FW01, BKF03}.
For $\nu_c^{ob} < \nu_X$ and assuming power law index $p=2$,
equations \ref{eq:nu_m} -- \ref{eq:F_nu} predict the flux at the X-ray
band after 11 hours to be $F_X^{ob} \simeq 0.3 \Z \dL \, \E \ee \, \mu
Jy$. This result depends only on the energy in the lepton component,
which is a fraction $\epsilon_e$ of the total explosion energy $E$.
For bursts with known luminosity distance, the rest-frame brightness is thus
\beq
F_{\nu,X} \simeq 3 \times 10^{26} \, \E \ee {\rm \, erg \, s^{-1} \, Hz^{-1}}
\eeq
(note that an order unity uncertainty may exist due to the
approximations used in deriving the equations, as well as the
assumption that $p=2$).  Comparison with the results presented in Table
4 and Figure \ref{xray_energy} shows that for a given X-ray brightness,
the energy emitted in $\gamma$ rays is similar to the energy carried
by the leptons. Thus, we can conclude that the $\gamma$ ray production
is efficient, $E_{\gamma,{\rm ISO}} / \epsilon_e E_{\rm ISO} \simeq 1$. This
result holds for at least the seven orders of magnitude in energies
considered in this work.

The similarity found between the X-ray fluxes of short and long
  GRBs implies that the high efficiency in $\gamma$ ray production is
  a property of both types. This result is consistent with the
  tentative conclusion of \citet{bpp+06}, who studied a single SGRB,
  namely GRB050509b.  

\section{Systematic biases and possible caveats}
\label{sec:statistics}

Our sample includes total of 37 short GRBs, out of which 16 ($43\%$)
have measured redshift. From those bursts, seven have measured optical and
X-ray fluxes at 11 hours and thus the fluence ratio at 11 hours (rest
frame) can be obtained, and nine have only upper limits on the observed
optical fluence, and thus only upper limits on the ratio at 11 hours
are known (see figures \ref{opt_energy}, \ref{fig:ratio}).  Therefore, a
question may arise as to whether our conclusions could be biased by
selection effects, and in particular, whether bursts in low density
regions might be excluded from our sample because their afterglows
would be expected to be too weak.

Consider, however, all bursts with $\gamma$-ray fluence larger than
$10^{-7} \rm{\, erg \, cm^{-2}}$ (see figure \ref{fig:opt_flux} and
tables \ref{short_table}, \ref{long_table}). Essentially all long GRBs
are in this group, as are more than half of the short bursts (22 out
of 37).  Furthermore, above this fluence, 8 out of 9 short bursts
which have reasonably deep ($< 3 \mu Jy$ or $AB=TBD$) mag early
searches, have identified optical counterparts.  Indeed, with one
exception, GRB~061210, all the optical upper limits lie above detections
with comparable $E_{\gamma}$.

If the standard model is correct and the prompt gamma-ray fluence
arises from an internal process, then choosing this sample in no way
selects bursts by the properties of their external medium.  However,
because an optical ID has been necessary to get a secure redshift for
a short GRB, this is the prime sample used to derive a relationship
between the fluxes $F_R$, $F_x$ and $E_{\gamma,{\rm ISO}}$ in Figures
\ref{opt_energy} -- \ref{fig:ratio}. Our fits then (in particular Figure \ref{opt_energy}) depend on a
sample which is essentially complete, and which does not depend upon
the properties of the external medium, unless the external medium
influences the observed $E_{\gamma}$.
% All of the upper limits seen in
%these figures arise from sources which are not part of this sample,
%with only one exception, namely GRB061210.

Of course, if $E_{\gamma,{\rm ISO}}$ depends upon the external medium, and in
particular its density, then our conclusions would be biased.  But
this bias only arises if perhaps the most interesting (and potentially
controversial) of our possible conclusions is true:  the initial
gamma-ray fluence is not entirely due to an internal process but is
also regulated by the medium external to the exploding object.

\begin{figure}
\plotone{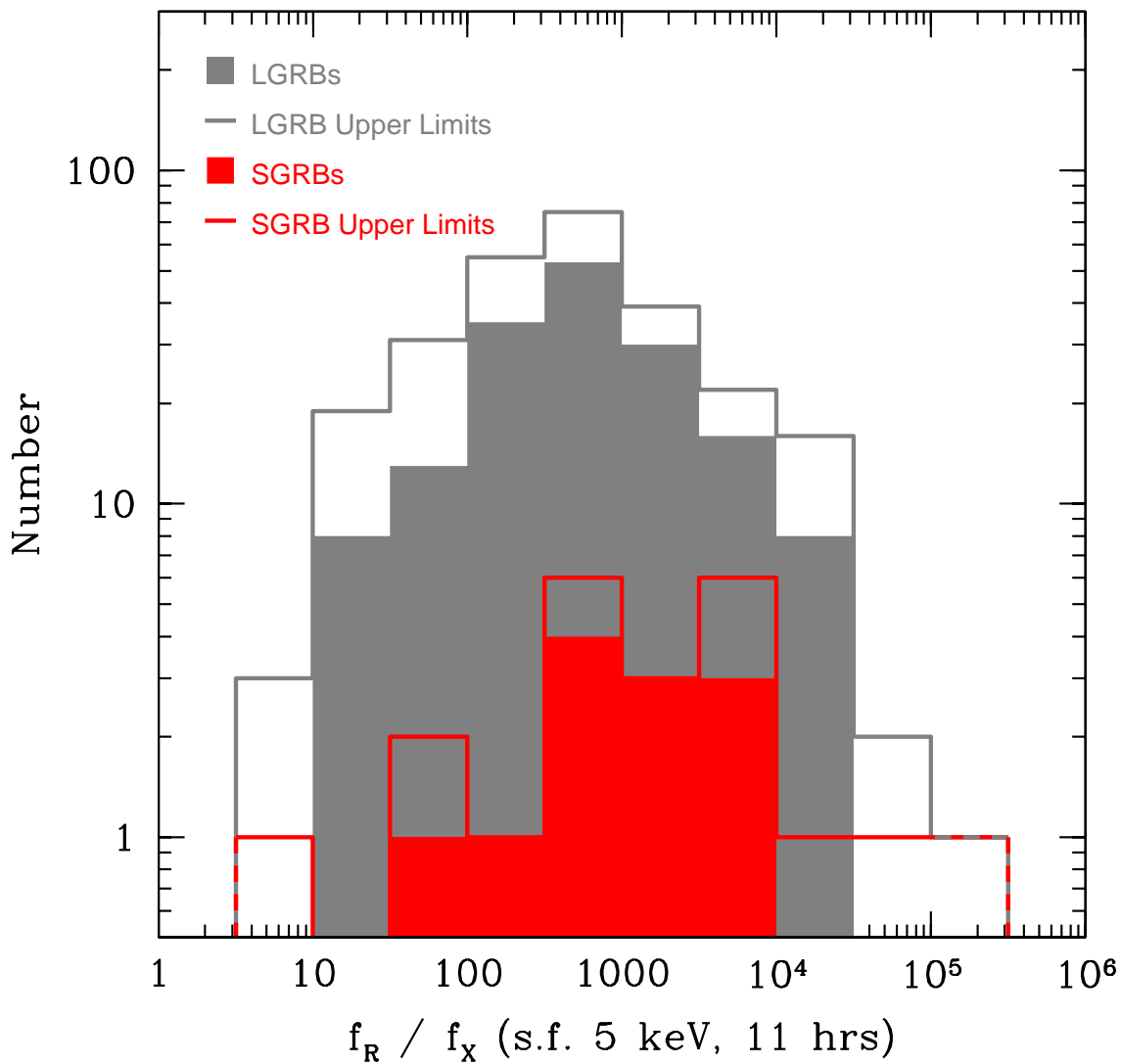}
\caption{Histogram of the ratio of the observed fluxes at the optical
  and X-band after 11 hours (observed time). Long GRBs are marked in
  gray, and short ones in red. When only upper limits are known, empty
boxes are drawn. Clearly, both samples show a nearly Gaussian shape in
a log-normal plot, with similar means.}
\label{fig:hist}
\end{figure}

A further comparison of optical and X-ray fluxes
is presented in figure \ref{fig:hist}. In this figure, we plot the
ratio of the observed fluxes for bursts,
using the full sample of bursts 
%that is presented in figures
%\ref{fig:opt_flux} and \ref{fig:X_flux} 
for which detections or upper
limits exist. Since the redshift, and thus $E_{\gamma,{\rm ISO}}$, is in most cases unknown, we
plot a histogram of the optical to X-ray flux
ratio at 11 hours observed time. The histogram clearly shows that the
two distributions exhibit a Gaussian-like behavior, with very similar
means, $<F_R/F_X> = 770$ for the long bursts and $<F_R/F_X> = 1130$
for the short bursts, where only those bursts with optical and X-ray
detections (no upper limits) were used in this calculation.  The
similar Gaussian shapes and similar means of the SGRB and LGRB
populations also suggests that contaminations, in particular incorrect 
identification of a long burst as short, does not significantly
affect the distribution (otherwise a bi-modal distribution would likely have
been seen - unless, of course both distributions have the same ratio,
which is the main conclusion here).   

Again, it is the fact that the values $<F_R/F_X>$ for the short bursts are near
1000 (and comparable or larger than those of long bursts) which is surprising.
It is then interesting that three very clear, well studied short bursts, 050709, 050724 and
051221A have values of  $<F_R/F_X>$ of about 7000, 10,000 and 500 respectively.
Examination of the table by the reader will show that even if some of the short bursts
in our sample are falsely identified short bursts, the bursts which are clearly
short agree with the claims presented here.

Additionally, it is possible that the ratio of  $f_R/f_X$ could be affected absorption of photons in
the host galaxy.   While many long GRBs show substantial 
X-ray columns \cite{crc+06, smp+07, gw01}, the large majority of the observed 10 keV
passband of the XRT is unaffected, and thus our X-ray fluxes should
be fairly accurate.    The measured optical absorption of long gamma-ray burst afterglows
often suggests lower column densities than the X-ray,  with typical optical absorptions  
in the range of one-tenth to one magnitude {\cite{smp+07, ckh+09}}.   
The effect of host absorption on short bursts has not yet been well studied.  But again,
the main danger would be a suppression of the optical relative to the X-ray.   However,
our main result is that the short bursts have brighter optical emission relative to
the X-ray than expected by the standard model and typically assumed
environmental densities.  
\section{Summary and discussion}
\label{sec:summary}

In this work, we present a comprehensive study of the optical and
X-ray afterglows obtained after 11 hours of 37 short and 421 long
GRBs, and compared the results to the energy emitted in $\gamma$-rays
during the prompt emission phase. This sample is the
largest used so far for this type of study.  We find a strong
correlation between the optical and X-ray afterglow brightnesses and
prompt $\gamma$-ray fluence of GRBs, a result similar to one reported
contemporaneously by 
by  \citet{g+08} \citet{Kann08}.  However, we also show that
the ratio of the optical to X-ray emission in the short burst
afterglows is comparable to that of the long-bursts.  This 
equivalence between burst types, and the absolute ratio we find can only be understood
in the framework of the standard theory if
 the circumburst density of SGRBs
is typically comparable to, or even higher than the circumburst density of
LGRBs. We point out that in a wind-like scenario for SGRBs can also explain 
this result if the average mass loss rate of the SGRB
progenitor is $\sim 30$ times lower than the characteristic mass loss
rate from a Wolf-Rayet star.  Finally, we have show
that in the framework of the standard model, in which the
  prompt emission fluence is independent of the environmental density,
  by selecting bursts with $\gamma$-ray fluence larger than $10^{-7}
  \rm{\, erg \, cm^{-2}}$ we obtain for short bursts above this fluence
  an essentially complete sample for analysis.

These results indicate that SGRB afterglows are not necessarily
dim due to a lower density in the circumburst environment.  While the
average SGRB afterglow is significantly dimmer than the average LGRB
afterglow, this appears  to be largely due to the lower total
energy release of the burst, as indicated by the fainter prompt
$\gamma$-ray emission.  We can thus conclude that if the prompt
emission is a result of internal shocks, then either the external
density of long and short bursts is similar, or the afterglow flux is
less sensitive to the density of the surrounding medium than
predicted.  On the other hand, if the prompt emission is due to an
external shock, both quantities the gamma-ray fluence and afterglow
intensity will depend on the external density and may scale
accordingly.  In this case we could attribute the faintness of {\it
  both} the prompt and afterglow emission to the external medium
density.

Due to the short lifetimes of the massive star progenitors of
long-duration GRBs, they have been traditionally expected to occur in
or near giant molecular clouds ($n \sim 10^{2.5}$ cm$^{-3}$;
\citealt{sal87}).  Some observational evidence supports this
hypothesis.  The hosts of LGRBs are typically blue and actively
forming massive stars \citep{ftm+99, sfc+01, ldm+03, chg04, gps+05}.
\citet{fls+06} find that LGRBs are preferentially located on the
brightest regions of these galaxies, which they interpret as evidence
that GRBs frequently occur on or near massive OB associations.  X-ray
observations of GRB afterglows show neutral hydrogen column densities
consistent those of Galactic giant molecular clouds \citep{gw01,
  jfl+06, smp+07}.  At optical wavelengths, moderate extinction
characteristic of dense regions is seen in some bursts \citep{kkz+07},
while in others the afterglow appears significantly extinguished
\citep{dfk+01, lfr+06, wfl+06, pdl+06, rvw+07}, and at high redshifts
where Ly$\alpha$ absorption can be found in the optical spectrum,
large hydrogen column densities (up to $10^{23}\, {\rm cm}^{-3}$) are
frequently found \citep[c.f.][]{jfl+07}.  Nonetheless, as noted
earlier, the density of the medium into which the relativistic flow of
the GRB expands may be dominated by an earlier wind from the
progenitor star rather than the ISM.

\begin{figure}

\plotone{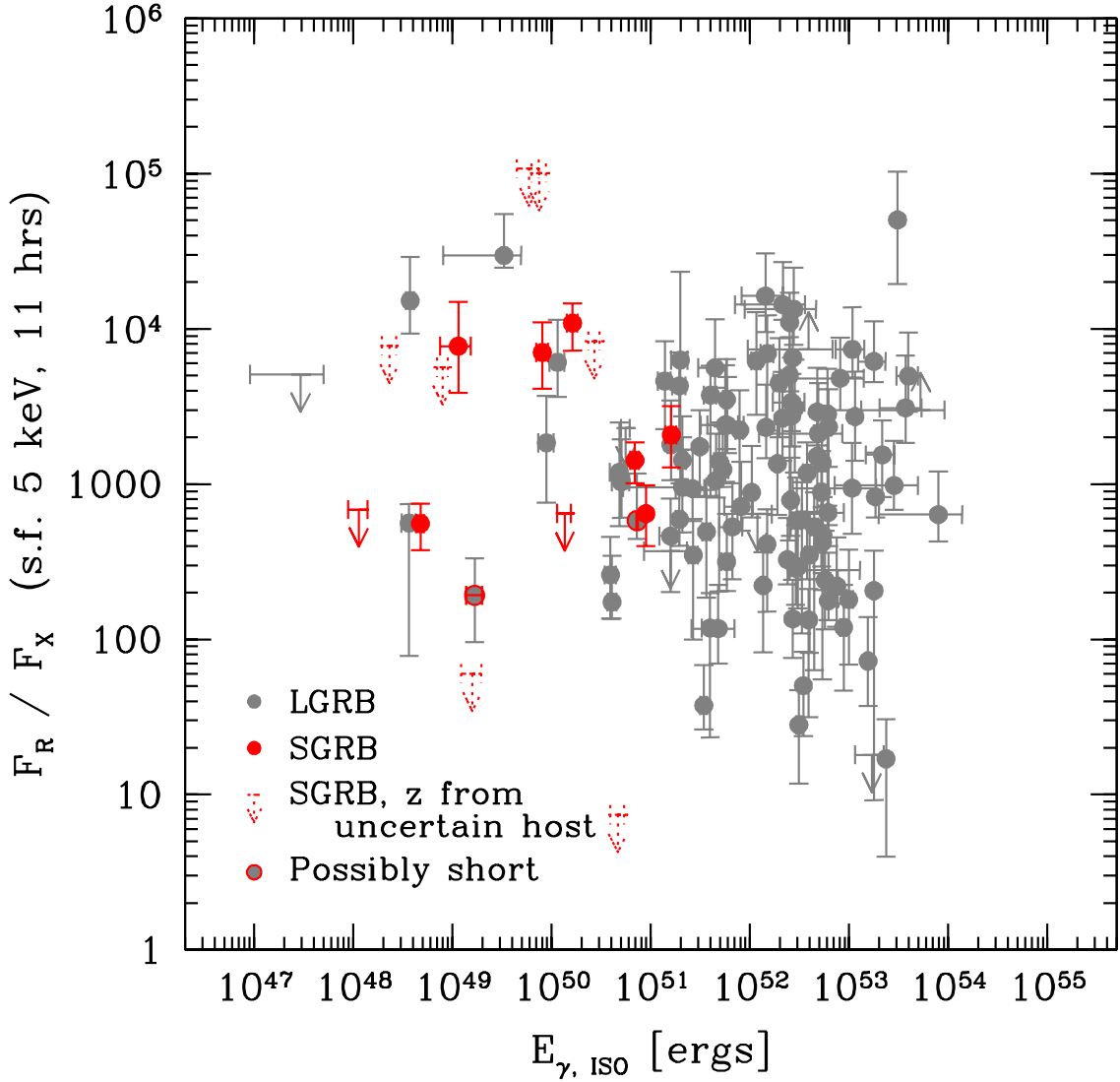}
\caption{The ratio of the total fluence in the optical over that in
  the x-ray versus the $\Egi$, measured at 11 hours (source frame).
  Once again the short and long bursts do not appear to differ, except
  in their typical $\Egi$.  There is no evidence in a suppression of
  this ratio either by a low density medium surrounding the short
  bursts or their lower typical $\Egi$, as would be expected from the
  standard theory if the synchroton frequency of the bursts were
  typically between the optical and X-ray.  However, the observed
  absolute values of this ratio are hard to explain by the standard
  theory if this is not the case.}
\label{fig:ratio}
\end{figure}

In contrast, SGRBs are found in both elliptical and star-forming
galaxies, a fact consistent with merger scenarios.  The ISM of
elliptical galaxies is sparse and has a typical density of only $n
\sim 10^{-2.5}$ cm$^{-3}$ \citep{fbp+06}.  \citet{Pan06}, however,
estimates a circumburst density of  $10^{-1} < n < 10^{3}
  \, {\rm cm}^{-3}$ for GRB 050724, a burst in an E/S0 galaxy \citep{bpc+05}.
  At the same time circumburst densities of $10^{-4} < n < 10^{-1} \, {\rm cm}^{-3}$ \cite{Pan06}
  and $10^{-3}$ to $10^{-1} \, {\rm cm}^{-3}$ \cite{sbk+06} are found
  for GRBs 050709 and 051221A, respectively, both of which occurred
  in apparently star-forming galaxies.
 As we have shown here, the lower ranges of these two latter
  densities  are inconsistent with the distribution of ratios of $F_X/F_R$ seen in
Figure \ref{fig:ratio}.  These
estimates of the SGRB circumburst density , however, assume a locally uniform
medium.  If the SGRBs were to typically form in a windlike structure,
this ratio would be acceptable.  Still, the common presence of a
wind structure around SGRBs would rule out the standard merger models.
The velocity by mass loss during the supernovae, let alone any kicks
by the supernovae, would easily move the compact remnants by more than
100 pc from their natal region before a binary merger occurred
\citep{bpb+06}. 

A potentially simple explanation of the results presented here is
 that both the afterglow and the prompt emission are due to an
external interaction \citep[e.g.,][]{Dermer08}. In this case, both the
afterglow and prompt emission might be expected to scale similarly.
However, it is difficult to explain the strong variability observed in
many bursts in such a model \citep[see, e.g.,][]{Mes06}. Furthermore,
observational evidence for a correlation between the density of a
circumburst medium and $\Egi$ is so far either weak or potentially
damaging.  \cite{tko+08} have suggested that SGRBs far from the center
of their host may be fainter than those closer in.  However, this work
uses a number of associations based on the brightest galaxy in a
rather large Swift XRT error circle, and therefore may in some cases
assign an incorrect offset, and thus will need to be confirmed on
larger, better samples.  In contrast, the only work to examine this
issue for long bursts has found a positive correlation between the
brightness of the burst and the distance of the burst from the center
of the host \citep{rlb02}.  This is, to first order, the opposite of
one would predict if the external medium played a role in the
gamma-ray emission.  It has, however, been noted that the bursts
GRBG 050509b and 050724, both of which were in early type
galaxies were intrinsically weaker than bursts
found in star-forming hosts \cite{Rhoads09}.  Nonetheless, we note again
that
the work by \citet{Pan06} suggests that the circumburst
medium around at GRB 0507024 was reasonably dense.

Several recent advances in the theory and modelling of the binary 
merger process and the local insterstellar medium could potentially explain 
our results.   \citet{vdh07} has pointed out that the large majority
of known double neutron-star binaries have low eccentricities,
suggesting that the second born neutron stars in binaries, unlike single neutron
stars, typically did not receive substantial ``kick'' velocities at birth.
Lower kicks will lead to lower systemic velocities, and a higher
probability that the system remains in a moderately dense ISM
when the binary merges.   Furthermore, the time to merger may often be relatively short.
\citet{bsf09} estimate that up to 70\% of NS-NS and NS-BH binaries
may merge with 100 Myr after formation.   Indeed, one burst that appears
to have likely been a short burst, GRB 060505, lies in an [HII] region on the outskirts
of its spiral host \citep{fwt+06,ocg+07,lk07}.   Some short bursts, however, 
occur in early-type galaxies.  Here recent result could also provide an
explanation for an apparent
ISM density of $1\, {\rm cm}^{-3}$.   Binary mergers in early-type galaxies
are almost certainly dominated by the evolving binaries in globular clusters \citep{gpm06},
and the most rapid evolution is likely to occur in the densest clusters.   These clusters however,
could have ionized gas densities of the required magnitude due to the colliding winds
of giants contained in the cluster \citep{prl09}.

Although we may not fully understand the mechanism,
we cannot escape the conclusion that SGRBs are fundamentally similar
in emission properties to LGRBs -- particularly the fainter LGRBs that
have typically been found at low redshift.  It has been suggested
that low-luminosity LGRBs (e.g. GRB 980425, GRB 031203 and GRB 060218)
have large, and perhaps nearly quasi-spherical, opening angles when
compared to traditional LGRBs \citep{skb+04, skn+06, lzv+07}.
Similarly, of the three SGRBs with estimated opening angles, GRB
050709 and GRB 050724 have large measured angles, $\Theta \sim 14^{o}$
\citep{ffp+05} and $\sim 25^{o}$ \citep{gbp+06, mcd+07} respectively,
while GRB 051221 has $4^o < \Theta < 8^o$ \citep{bgc+06, sbk+06},
which is more typical of traditional high $\Egi$ LGRB opening angles.
Yet both long and short populations appear to largely fall on a simple
log-linear relationship between afterglow intensity and gamma-ray
luminosity over six orders of magnitude in energy.  Within this
relationship there may be strong hints as to the nature of the
progenitors of SGRBs, or perhaps a new more fundamental understanding
of the GRB emission mechanism.

%% ACKNOWLEDGMENTS

\acknowledgments

This work made use of data supplied by the UK {\it Swift} Science Data
Centre at the University of Leicester.  We gratefully acknowledge
J. Racusin, N. Gehrels, D. Burrows, and the {\it Swift} team for
generously sharing pre-published XRT values.  We also thank
A. Levan, N. Gehrels, E, Waxman, K. Belczynski, E. Ramirez-Ruiz, E. van den Heuvel,
R. Sari and J. Graham for 
helpful discussions, J. Greiner for
his online GRB Table and R. Quimby, E. McMahon and J. Murphy for
 the GRBlog.  AP wishes to acknowledge the support of the
Riccardo Giacconi fellowship award of the Space Telescope Science
Institute.

%% BIBLIOGRAPHY

\clearpage

%% FIGURES

%\begin{figure}
%Figure 17
%\epsscale{.80}
%\plotone{grb_xE_2.pdf}
%\caption{
%}
%\label{grb_xE_2}
%\end{figure}

%% TABLES

\clearpage

\begin{landscape}

\small

\begin{longtable}{p{0.45in}ccccccccccc}
\caption{Observed Prompt and Afterglow Properties of Short-Duration GRBs} \\
\hline \hline
& & & & Fluence & Log & 11 hr & 11 hr 0.2--10 & Log \\
GRB & Satellite & Channel & z & [10$^{-7}$ & E$_{\gamma, ISO}$ & $R$-band & keV [10$^{-14}$ & F$_{R}$/ & Ref. \\
& & [keV] &  &  erg cm$^{-2}$] & [erg] & [$\mu$Jy] & erg cm$^{-2}$ s$^{-1}$] & F$_{X}$ & \\
\endfirsthead
\multicolumn{3}{c}{{\tablename} \thetable{} -- Continued} \\[0.5ex]
\hline \hline
& & & & Fluence & Log & 11 hr & 11 hr 0.2--10 & Log \\
GRB & Satellite & Channel & z & [10$^{-7}$ & E$_{\gamma, ISO}$ & $R$-band & keV [10$^{-14}$ & F$_{R}$/ & Ref. \\
& & [keV] &  &  erg cm$^{-2}$] & [erg] & [$\mu$Jy] & erg cm$^{-2}$ s$^{-1}$] & F$_{X}$ & \\
 \\[-1.8ex]
\endhead
\multicolumn{3}{l}{{Continued on Next Page\ldots}} \\
\endfoot
\\[-2ex] \hline \hline \\[0.6ex]
{\parbox{8.3in}{\footnotesize 
$^\dagger$: probable redshift, $^{C}$: cluster redshift, $^{a}$ Uses estimated N$_{H}$, 1. \citet{hcm+00c},
2. \citet{cag00}, 3. \citet{hcf+00}, 4. \citet{pas00c}, 5. \citet{hcm+01}, 6. \citet{pmb01}, 7. \citet{dls+06}, 8. \citet{pas+01}, 9. \citet{slk+05}, 10. \citet{lsv+02}, 11. \citet{hcm+02}, 12. \citet{psa02b}, 13. \citet{hcm+02b}, 14. \citet{nmj+02}, 15. \citet{sbb+08}, 16. \citet{cbk+05}, 17. \citet{gso+05}, 18. \citet{bpp+06}, 19. \citet{vil+05},
20. \citet{ffp+05}, 21. \citet{g+08}, 22. \citet{mcd+07}, 23. \citet{bpc+05}, 24. \citet{ebp+07}, 25. \citet{bgs+05}, 26. \citet{b05}, 27. \citet{ltj+08}, 28. \citet{plb+05}, 29. \citet{htt+05}, 30. \citet{mtm+05}, 31. \citet{bb05}, 32. \citet{bf05}, 33. \citet{cbh+05}, 34. \citet{gms+05}, 35. \citet{sbk+06}, 36. \citet{bfp+07},  37. \citet{ltf+06}, 38. \citet{pbf+06b}, 39. \citet{bpc+06}, 40. \citet{pcm+06b}, 41. \citet{cfb+06}, 42. \citet{d+08}, 43. \citet{sdp+07b}, 44. \citet{b07}, 45. \citet{cfp06}, 46. \citet{c06}, 47. \citet{b06}, 48. \citet{spc+07}, 49. \citet{jrh+07}, 50. \citet{mtc+07}, 51. \citet{zzq+07}, 52. \citet{hdm07}, 53. \citet{pbm+07}, 54. \citet{gam+07b},
55. \citet{pvd+07}, 56. \citet{bbc+07b}, 57. \citet{ltb+07}, 58. \citet{gfl+07}
%59. \citet{zbp+07}
%60. \citet{crb+07}, 61. \citet{cfc+07b}, 62. \citet{gsc+07}, 63. \citet{bk07}, 64. \citet{mbb+07d}, 65. \citet{ptb07}, 66. \citet{mbb2+07}, 67. \citet{sop+07}, 68. \citet{xzq+07}, 69. \citet{psf+07}, 70. \citet{kb07}, 71. \citet{sbb+07c}, 72. \citet{w+08}, 73. \citet{bmr07}
}}
\endlastfoot
001025B & {\it \small Ulysses} & 25--100 &  & 2.0 & & $<382.6$ &   & & {\footnotesize 1,2} \\
001204 & {\it \small Ulysses} & 25--100 &  & 8.6 & & $<190.9$ &   & & {\footnotesize 3,4} \\
010119 & {\it \small Ulysses} & 25--100 &  & 3.2 & & $<1970.3$ &   & & {\footnotesize 5,6} \\
010326B & {\it \small HETE} & 30--400 &  & 3.3$\pm$0.8 & & $<26.0$ &   & & {\footnotesize 7,8}\\
020531 & {\it \small HETE} & 30--400 &  & 11.1$^{+1.4}_{-1.3}$ & & $<16.9$  & & & {\footnotesize 9,10}\\
020603 & {\it \small Ulysses} & 25--100 &  & 60.0 & & $<708.7$ &   & & {\footnotesize 11,12}\\
021201 & {\it \small Ulysses} & 25--100 &  & 2.0 & & $<58.0$ &   & & {\footnotesize 13,14}\\
050202 & {\it Swift} & 15--150 &  & 0.3$\pm$0.1 & & $<55.3$ &   & & {\footnotesize 15, 16}\\
\end{longtable}
\end{landscape}

{\footnotesize
59. \citet{zbp+07}
60. \citet{crb+07}, 61. \citet{cfc+07b}, 62. \citet{gsc+07}, 63. \citet{bk07}, 64. \citet{mbb+07d}, 65. \citet{ptb07}, 66. \citet{mbb2+07}, 67. \citet{sop+07}, 68. \citet{xzq+07}, 69. \citet{psf+07}, 70. \citet{kb07}, 71. \citet{sbb+07c}, 72. \citet{w+08}, 73. \citet{bmr07}
}

\clearpage

\begin{landscape}

\small
\begin{center}
\begin{longtable}{p{0.45in}ccccccccccc}
\caption{Observed Prompt and Afterglow Properties of Long-Duration GRBs} \\
\hline \hline
& & & & Fluence & Log & 11 hr & 11 hr 0.2--10 & F$_{R}$/ \\
GRB & Satellite & Channel & z & [10$^{-7}$ & E$_{ISO}$ & $R$-band & keV [10$^{-14}$ & F$_{X}$ & Ref. \\
& & [keV] &  &  erg cm$^{-2}$] & [erg] & [$\mu$Jy] & erg cm$^{-2}$ s$^{-1}$] & & \\
\endfirsthead
\multicolumn{3}{c}{{\tablename} \thetable{} -- Continued} \\[0.5ex]
\hline \hline
& & & & Fluence & Log & 11 hr & 11 hr 0.2--10 & F$_{R}$/ \\
GRB & Satellite & Channel & z & [10$^{-7}$ & E$_{ISO}$ & $R$-band & keV [10$^{-14}$ & F$_{X}$ & Ref. \\
& & [keV] &  &  erg cm$^{-2}$] & [erg] & [$\mu$Jy] & erg cm$^{-2}$ s$^{-1}$] & & \\
 \\[-1.8ex]
\endhead
\multicolumn{3}{l}{{Continued on Next Page\ldots}} \\
\endfoot
\\[-3ex] \hline \hline \\[0.1ex]
{\parbox{8.5in}{\footnotesize $^{a}$: 1.6--10.0 keV, $^{b}$: 2--10 keV, $^{c}$: 0.5--6.0 keV, $^{d}$: Uses estimated N$_{H}$, 1. \citet{dpg+06} 2. \citet{cgh+97}  3. \citet{ggv+97}, 4. \citet{dkb+99}, 5. \citet{ggv+97c}, 6. \citet{dgm+97}, 7. \citet{mdk+97}, 8. \citet{pmp+99}, 9. \citet{ckp+97}, 10. \citet{gcp+99}, 11. \citet{mhg+05}, 12. \citet{sdh+04}, 13. \citet{ggv+98}, 14. \citet{muy+97}, 15. \citet{dfk+01}, 16. \citet{ddc+98}, 17. \citet{afa+99}, 18. \citet{ggv+97b}, 19. \citet{mt98}, 20. \citet{ggv+98b}, 21. \citet{rlm+99}, 22. \citet{tsc+98}, 23. \citet{gvg+98}, 24. \citet{vhc+00}, 25. \citet{dbk03}, 26. \citet{htn+02}, 27. \citet{dkg+98}, 28. \citet{vgo+99} 
}}
\endlastfoot
970111 & {\it \small BeppoSAX} & 40--700 &  & 430$\pm$30 & & $<$33.6 & 7.5$\pm4.7^{a}$ & & {\scriptsize 1, 2} \\
970228 & {\it \small BeppoSAX} & 40--700	& 0.695 & 64.5 & 51.91$^{+0.12}_{-0.17}$ & 22.0$^{+4.0}_{-3.5	}$ & 208.0$\pm$ $27.0^{a}$ & 2.86$^{+0.29}_{-0.15}$ & {\scriptsize 1, 3, 4} \\
970402 & {\it \small BeppoSAX} & 40--700	& & 82.0$\pm$9.0 & &  $<$44.0 & 13.5$\pm7.3^{a}$ & & {\scriptsize 1, 5} \\
970508 & {\it \small BeppoSAX} & 40--700	& 0.835 & 14.5 & 51.42$^{+0.12}_{-0.17}$  & 6.5$^{+1.6}_{-1.3}$ & 57.2$\pm9.0^{a}$ & 2.97$^{+0.25}_{-0.21}$ & {\scriptsize 1, 6, 7} \\
970616 & {\it \small BATSE} & 20--100 & & 109.7$\pm$0.6	& & $<$14.0 & 385.8$^{+208.3b,d}_{-186.8}$ & & {\scriptsize 8, 9, 10} \\
970815 & {\it \small BATSE} & 20--100 & & 52.5$\pm$0.6 & & 14.2$^{+12.3}_{-6.7}$ &	 498.9$^{+471.8b}_{-286.4}$ & & {\scriptsize 8, 11, 12} \\
970828 & {\it \small BATSE} & 20--100 & 0.958 & 700.0 & 53.23$^{+0.12}_{-0.17}$ & $<$0.6 & 198.8$^{+113.5b,d}_{-91.1}$ & $<$1.26 & {\scriptsize 13, 14, 15} \\
971214 & {\it \small BeppoSAX} & 40--700 & 3.42 & 64.9 & 53.03$^{+0.24}_{-0.60}$ & 5.6$^{+0.5}_{-0.4}$ & 63.5$\pm9.1^{a}$ & 2.97$\pm$0.29 & {\scriptsize 1, 16} \\
\end{longtable}
\end{center}
\end{landscape}

%\begin{table}
%\begin{center}
%\caption{Best-fit Results of the Observed Properties of Long and Short GRBs}
%\begin{tabular}{lccccc}
%\\
%\hline\hline
%& slope & zero-point & $\chi^2/dof$ & zero-point, & $\chi^2/dof$ \\
%& $\alpha$ & $a$ & & w/ $\alpha = 1$ & \\
%Short, optical & 1.63$\pm$0.91 & 10.66$\pm$7.52 & 0.65 & 6.74$\pm$0.17 & 0.73 \\
%Long, optical & 1.00$\pm$0.11 & 6.75$\pm$0.63 & 1.31 & 6.77$\pm$0.05 & 1.31 \\
%Short, X-ray & 0.80$\pm$0.25 & -7.70$\pm$1.68 & 0.61 & -6.39$\pm$0.15 & 0.66 \\
%Long, X-ray & 1.01$\pm$0.08 & -6.33$\pm$0.48 & 0.90 & -6.36$\pm$0.04 & 0.90 \\
%\hline
%\end{tabular}
%\end{center}
%\label{parameters}
%\end{table}

\begin{table}
\begin{center}
\caption{Best-fit Results of the Observed Properties of Long and Short GRBs}
\begin{tabular}{lccccc}
\\
\hline\hline
& slope & zero-point & $\chi^2/dof$ & zero-point, & $\chi^2/dof$ \\
& $\alpha$ & $a$ & & w/ $\alpha = 1$ & \\
Short, optical & 1.63$\pm$0.91 & 10.66$\pm$7.52 & 0.65 & 6.74$\pm$0.17 & 0.73 \\
Long, optical & 0.97$\pm$0.11 & 6.58$\pm$0.62 & 1.30 & 6.78$\pm$0.05 & 1.30  \\
Short, X-ray & 1.06$\pm$0.30 & -5.94$\pm$2.02 & 0.67 & -6.33$\pm$0.16 & 0.66 \\
Long, X-ray & 1.10$\pm$0.09 & -5.59$\pm$0.54 & 1.05 & -6.18$\pm$0.04 & 1.06 \\
\hline
\end{tabular}
\end{center}
\label{parameters}
\end{table}

%\begin{table}
%\begin{center}
%\caption{Best-fit Results of the Rest-Frame Properties of Long and Short GRBs}
%\begin{tabular}{lccccc}
%\\
%\hline\hline
%& slope & zero-point & $\chi^2/dof$ & zero-point, & $\chi^2/dof$ \\
%& $\alpha$ & $a$ & & w/ $\alpha = 1$ & \\
%Short, optical & 1.05$\pm$0.26 & -24.41$\pm$13.19 & 0.31 & -21.90$\pm$0.22 & 0.31 \\
%Long, optical & 0.93$\pm$0.08 & -18.22$\pm$4.29 & 0.90 & -21.68$\pm$0.07 & 0.90 \\
%Short, X-ray & 0.88$\pm$0.20 & -18.91$\pm$10.12 & 0.52 & -25.17$\pm$0.20 & 0.54 \\
%Long, X-ray & 1.08$\pm$0.07 & -29.22$\pm$3.90 & 1.08 & -24.88$\pm$0.07 & 1.08 \\
%\hline
%\end{tabular}
%\end{center}
%\label{parameters2}
%\end{table}

\begin{table}
\begin{center}
\caption{Best-fit Results of the Rest-Frame Properties of Long and Short GRBs}
\begin{tabular}{lccccc}
\\
\hline\hline
& slope & zero-point & $\chi^2/dof$ & zero-point, & $\chi^2/dof$ \\
& $\alpha$ & $a$ & & w/ $\alpha = 1$ & \\
Short, optical & 1.05$\pm$0.26 & -24.42$\pm$13.20 & 0.31 & -21.90$\pm$0.22 & 0.31 \\
Long, optical &0.93$\pm$0.09 & -19.20$\pm$4.44 & 0.90 & -21.68$\pm$0.07 & 0.90 \\
Short, X-ray & 1.00$\pm$0.21 & -25.18$\pm$10.54 & 0.71 & -25.04$\pm$0.19 & 0.71 \\
Long, X-ray & 1.05$\pm$0.07 & -27.32$\pm$3.73 & 1.09 & -24.74$\pm$0.07 & 1.10 \\
\hline
\end{tabular}
\end{center}
\label{parameters2}
\end{table}

\end{document}

\begin{landscape}

\small

\begin{longtable}{p{0.5in}ccccccccccc}
\caption{Observed Prompt and Afterglow Properties of Short-Duration GRBs} \\
\hline \hline
& & & & Fluence & Log & 11 hr & 11 hr 0.2--10 & Log \\
GRB & Satellite & Channel & z & [10$^{-7}$ & E$_{\gamma, ISO}$ & $R$-band & keV [10$^{-14}$ & F$_{R}$/ & Ref. \\
& & [keV] &  &  erg cm$^{-2}$] & [erg] & [$\mu$Jy] & erg cm$^{-2}$ s$^{-1}$] & F$_{X}$ & \\
\endfirsthead
\multicolumn{3}{c}{{\tablename} \thetable{} -- Continued} \\[0.5ex]
\hline \hline
& & & & Fluence & Log & 11 hr & 11 hr 0.2--10 & Log \\
GRB & Satellite & Channel & z & [10$^{-7}$ & E$_{\gamma, ISO}$ & $R$-band & keV [10$^{-14}$ & F$_{R}$/ & Ref. \\
& & [keV] &  &  erg cm$^{-2}$] & [erg] & [$\mu$Jy] & erg cm$^{-2}$ s$^{-1}$] & F$_{X}$ & \\
 \\[-1.8ex]
\endhead
\multicolumn{3}{l}{{Continued on Next Page\ldots}} \\
\endfoot
\\[-2ex] \hline \hline \\[0.6ex]
{\parbox{8in}{\footnotesize $^\dagger$: probable redshift, $^{C}$: cluster redshift, $^{a}$ Uses estimated N$_{H}$, 1. \citet{hcm+00c},
2. \citet{cag00}, 3. \citet{hcf+00}, 4. \citet{pas00c}, 5. \citet{hcm+01}, 6. \citet{pmb01}, 7. \citet{dls+06}, 8. \citet{pas+01}, 9. \citet{slk+05}, 10. \citet{lsv+02}, 11. \citet{hcm+02}, 12. \citet{psa02b}, 13. \citet{hcm+02b}, 14. \citet{nmj+02}, 15. \citet{sbb+08}, 16. \citet{cbk+05}, 17. \citet{gso+05}, 18. \citet{bpp+06}, 19. \citet{vil+05},
20. \citet{ffp+05}, 21. \citet{g+08}, 22. \citet{mcd+07}, 23. \citet{bpc+05}, 24. \citet{ebp+07}, 25. \citet{bgs+05}, 26. \citet{b05}, 27. \citet{ltj+08}, 28. \citet{plb+05}, 29. \citet{htt+05}, 30. \citet{mtm+05}, 31. \citet{bb05}, 32. \citet{bf05}, 33. \citet{cbh+05}, 34. \citet{gms+05}, 35. \citet{sbk+06}, 36. \citet{bfp+07},  37. \citet{ltf+06}, 38. \citet{pbf+06b}, 39. \citet{bpc+06}, 40. \citet{pcm+06b}, 41. \citet{cfb+06}, 42. \citet{d+08}, 43. \citet{sdp+07b}, 44. \citet{b07}, 45. \citet{cfp06}, 46. \citet{c06}, 47. \citet{b06}, 48. \citet{spc+07}, 49. \citet{jrh+07}, 50. \citet{mtc+07}, 51. \citet{zzq+07}, 52. \citet{hdm07}, 53. \citet{pbm+07}, 54. \citet{gam+07b},
55. \citet{pvd+07}, 56. \citet{bbc+07b}, 57. \citet{ltb+07}, 58. \citet{gfl+07}
%59. \citet{zbp+07}
%60. \citet{crb+07}, 61. \citet{cfc+07b}, 62. \citet{gsc+07}, 63. \citet{bk07}, 64. \citet{mbb+07d}, 65. \citet{ptb07}, 66. \citet{mbb2+07}, 67. \citet{sop+07}, 68. \citet{xzq+07}, 69. \citet{psf+07}, 70. \citet{kb07}, 71. \citet{sbb+07c}, 72. \citet{w+08}, 73. \citet{bmr07}
}}
\endlastfoot
001025B & {\it \small Ulysses} & 25--100 &  & 2.0 & & $<382.6$ &   & & {\footnotesize 1,2} \\
001204 & {\it \small Ulysses} & 25--100 &  & 8.6 & & $<190.9$ &   & & {\footnotesize 3,4} \\
010119 & {\it \small Ulysses} & 25--100 &  & 3.2 & & $<1970.3$ &   & & {\footnotesize 5,6} \\
010326B & {\it \small HETE} & 30--400 &  & 3.3$\pm$0.8 & & $<26.0$ &   & & {\footnotesize 7,8}\\
020531 & {\it \small HETE} & 30--400 &  & 11.1$^{+1.4}_{-1.3}$ & & $<16.9$  & & & {\footnotesize 9,10}\\
020603 & {\it \small Ulysses} & 25--100 &  & 60.0 & & $<708.7$ &   & & {\footnotesize 11,12}\\
021201 & {\it \small Ulysses} & 25--100 &  & 2.0 & & $<58.0$ &   & & {\footnotesize 13,14}\\
050202 & {\it Swift} & 15--150 &  & 0.3$\pm$0.1 & & $<55.3$ &   & & {\footnotesize 15, 16}\\
\end{longtable}
\end{landscape}

\clearpage

{\footnotesize \noindent 59. \citet{zbp+07}, 60. \citet{crb+07}, 61. \citet{cfc+07b}, 62. \citet{gsc+07}, 63. \citet{bk07}, 64. \citet{mbb+07d}, 65. \citet{ptb07}, 66. \citet{mbb2+07}, 67. \citet{sop+07}, 68. \citet{xzq+07}, 69. \citet{psf+07}, 70. \citet{kb07}, 71. \citet{sbb+07c}, 72. \citet{w+08}, 73. \citet{bmr07}
}

\clearpage

\begin{landscape}

\small
\begin{longtable}{p{0.5in}ccccccccccc}

\caption{Observed Prompt and Afterglow Properties of Long-Duration GRBs (con't)} \\
\hline \hline
& & & & Fluence & Log & 11 hr & 11 hr 0.2--10 & Log \\
GRB & Satellite & Channel & z & [10$^{-7}$ & E$_{\gamma, ISO}$ & $R$-band & keV [10$^{-14}$ & F$_{R}$/ & Ref. \\
& & [keV] &  &  erg cm$^{-2}$] & [erg] & [$\mu$Jy] & erg cm$^{-2}$ s$^{-1}$] & F$_{X}$ & \\
\endfirsthead
\multicolumn{3}{c}{{\tablename} \thetable{} -- Continued} \\[0.5ex]
\hline \hline
& & & & Fluence & Log & 11 hr & 11 hr 0.2--10 & Log \\
GRB & Satellite & Channel & z & [10$^{-7}$ & E$_{\gamma, ISO}$ & $R$-band & keV [10$^{-14}$ & F$_{R}$/ & Ref. \\
& & [keV] &  &  erg cm$^{-2}$] & [erg] & [$\mu$Jy] & erg cm$^{-2}$ s$^{-1}$] & F$_{X}$ & \\
 \\[-1.8ex]
\endhead
\multicolumn{3}{l}{{Continued on Next Page\ldots}} \\
\endfoot
\\[-2ex] \hline \hline \\[0.6ex]
{\parbox{9in}{\footnotesize $^{a}$: 1.6--10.0 keV, $^{b}$: 2--10 keV, $^{c}$: 0.5--6.0 keV, $^{d}$: Uses estimated N$_{H}$, 1. \citet{dpg+06} 2. \citet{cgh+97}  3. \citet{ggv+97}, 4. \citet{dkb+99}, 5. \citet{ggv+97c}, 6. \citet{dgm+97}, 7. \citet{mdk+97}, 8. \citet{pmp+99}, 9. \citet{ckp+97}, 10. \citet{gcp+99}, 11. \citet{mhg+05}, 12. \citet{sdh+04}, 13. \citet{ggv+98}, 14. \citet{muy+97}, 15. \citet{dfk+01}, 16. \citet{ddc+98}, 17. \citet{afa+99}, 18. \citet{ggv+97b}, 19. \citet{mt98}, 20. \citet{ggv+98b}, 21. \citet{rlm+99}, 22. \citet{tsc+98}, 23. \citet{gvg+98}, 24. \citet{vhc+00}, 25. \citet{dbk03}, 26. \citet{htn+02}, 27. \citet{dkg+98}, 28. \citet{vgo+99} 29. \citet{faf+98}, 30. \citet{plh+99}, 31. \citet{cgd+98}, 32. \citet{kdo+99}, 33. \citet{pmp+99b}, 34. \citet{ssh+99}, 35. \citet{bbk+03}, 36. \citet{mpp+99}, 37. \citet{vgr+99}, 38. \citet{hbf+99}, 39. \citet{kpg+99}, 40. \citet{cgp+99}, 41. \citet{hc99}, 42. \citet{phj+99}, 43. \citet{jhp+99}, 44. \citet{ldm+02}, 45. \citet{mpp+00}, 46. \citet{gvr+99}, 47. \citet{svb+00}, 48. \citet{gpe+99}, 49. \citet{tuh+99}, 50. \citet{ppm+99}, 51. \citet{hcm+00}, 52. \citet{das+99}, 53. \citet{csg+01}, 54. \citet{gcp06}, 55. \citet{vrh+99}, 56. \citet{hum+00}, 57. \citet{gvh+00}, 58. \citet{hcm00}, 59. \citet{kjs+00}, 60. \citet{hcm00b}, 61. \citet{ahp+00}, 62. \citet{pfg+00}, 63. \citet{rws+00}, 64. \citet{jfg+01}, 65. \citet{cdd+00}, 66. \citet{rf01}, 67. \citet{hck+00}, 68. \citet{pjg+00}, 69. \citet{hkb+00}, 70. \citet{hcc00}, 71. \citet{hbk+00}, 72. \citet{ppw+00}, 73. \citet{kw00}, 74. \citet{pas00}, 75. \citet{hcm00c}, 76. \citet{ksm+00}, 77. \citet{umh+00}, 78. \citet{h00}, 79. \citet{jph+00}, 80. \citet{jph+00b}, 81. \citet{pmp+00}, 82. \citet{hcm+00d}, 83. \citet{dc00}, 84. \citet{sgj+00}, 85. \citet{hcm+00b}, 86. \citet{bgg+00}, 87. \citet{hcm+00z} 88. \citet{gpg+00}, 89. \citet{hcm+00e}, 90. \citet{fjg+01}, 91. \citet{hgm+00}, 92. \citet{pmp+00b}, 93. \citet{hcm+00f}, 94. \citet{mpp+00b}, 95. \citet{hcm+00g}, 96. \citet{pas00b}, 97. \citet{hcm+00h}, 98. \citet{pbk+02}, 99. \citet{hmg+00}, 100. \citet{fgd+01}, 101. \citet{hcm+00i}, 102. \citet{ccg+02}, 103. \citet{hcm+00j}, 104. \citet{hkh+00}, 105. \citet{hcs00}, 106. \citet{phh+06}, 107. \citet{vrp+00}, 
}}
\endlastfoot
970111 & {\it \small BeppoSAX} & 40--700 &  & 430$\pm$30 & & $<$33.6 & 7.5$\pm4.7^{a}$ & & {\scriptsize 1, 2} \\
970228 & {\it \small BeppoSAX} & 40--700	& 0.695 & 64.5 & 51.91$^{+0.12}_{-0.17}$ & 22.0$^{+4.0}_{-3.5	}$ & 208.0$\pm$ $27.0^{a}$ & 2.86$^{+0.29}_{-0.15}$ & {\scriptsize 1, 3, 4} \\
970402 & {\it \small BeppoSAX} & 40--700	& & 82.0$\pm$9.0 & &  $<$44.0 & 13.5$\pm7.3^{a}$ & & {\scriptsize 1, 5} \\
970508 & {\it \small BeppoSAX} & 40--700	& 0.835 & 14.5 & 51.42$^{+0.12}_{-0.17}$  & 6.5$^{+1.6}_{-1.3}$ & 57.2$\pm9.0^{a}$ & 2.97$^{+0.25}_{-0.21}$ & {\scriptsize 1, 6, 7} \\
970616 & {\it \small BATSE} & 20--100 & & 109.7$\pm$0.6	& & $<$14.0 & 385.8$^{+208.3b,d}_{-186.8}$ & & {\scriptsize 8, 9, 10} \\
970815 & {\it \small BATSE} & 20--100 & & 52.5$\pm$0.6 & & 14.2$^{+12.3}_{-6.7}$ &	 498.9$^{+471.8b}_{-286.4}$ & & {\scriptsize 8, 11, 12} \\
970828 & {\it \small BATSE} & 20--100 & 0.958 & 700.0 & 53.23$^{+0.12}_{-0.17}$ & $<$0.6 & 198.8$^{+113.5b,d}_{-91.1}$ & $<$1.26 & {\scriptsize 13, 14, 15} \\
971214 & {\it \small BeppoSAX} & 40--700 & 3.42 & 64.9 & 53.03$^{+0.24}_{-0.60}$ & 5.6$^{+0.5}_{-0.4}$ & 63.5$\pm9.1^{a}$ & 2.97$\pm$0.29 & {\scriptsize 1, 16} \\
\end{longtable}
\end{landscape}

{\footnotesize \noindent
108. \citet{hcm+00k}, 109. \citet{pms+00}, 110. \citet{hcm+00l}, 111. \citet{z00}, 112. \citet{bol+03}, 113. \citet{mpp+01}, 114. \citet{slk+05}, 115. \citet{z01}, 116. \citet{rsk+01}, 117. \citet{jpg+01}, 118. \citet{grb+03}, 119. \citet{hgm+01}, 120. \citet{ohp+01}, 121. \citet{pas+01b}, 122. \citet{hgm+01b}, 123. \citet{pgy+01}, 124. \citet{ajh+01}, 125. \citet{pkb+02}, 126. \citet{amr+01}, 127. \citet{hik+01}, 128. \citet{mpp+01b}, 129. \citet{igs+01}, 130. \citet{gks+03}, 131. \citet{ltl+01}, 132. \citet{vsf+06}, 133. \citet{hsg+02}, 134. \citet{scg+01}, 135. \citet{hmg+03}, 136. \citet{bkb+02}, 137. \citet{bfk+07}, 138. \citet{cgg+02b}, 139. \citet{llt+02}, 140. \citet{pr02}, 141. \citet{pjp02}, 142. \citet{mdb+02}, 143. \citet{hcf+02}, 144. \citet{mpp+03}, 145. \citet{lnf+05}, 146. \citet{cgc+02}, 147. \citet{hfg+02}, 148. \citet{psa02}, 149. \citet{afi+04}, 150. \citet{hmg+02}, 151. \citet{hn02}, 152. \citet{hcm+02c}, 153. \citet{ghj+02}, 154. \citet{hcm+02d}, 155. \citet{gcc+02}, 156. \citet{hcm+02e}, 157. \citet{gcc+02b}, 158. \citet{kg02}, 159. \citet{pbb+02}, 160. \citet{pbg+02}, 161. \citet{ls03}, 162. \citet{jff+05}, 163. \citet{lbf+02}, 164. \citet{psa02c}, 165. \citet{snc+07}, 166. \citet{srw+02}, 167. \citet{hwf+03}, 168. \citet{hmg+02b}, 169. \citet{ckw+02}, 170. \citet{ssh+02}, 171. \citet{hcm+02f}, 172. \citet{bf02}, 173. \citet{fp02}, 174. \citet{ssc+02}, 175. \citet{ysg+02}, 176. \citet{nrs02}, 177. \citet{vfh+03}, 178. \citet{nrp+06}, 179. \citet{mgb+03}, 180. \citet{cga+02}, 181. \citet{lfr+06}, 182. \citet{gmh+03}, 183. \citet{fph+03}, 184. \citet{hcm+03}, 185. \citet{nhl+03}, 186. \citet{kgr+04}, 187. \citet{mgt+03}, 188. \citet{cgg+03}, 189. \citet{vel+04}, 190. \citet{gkh03}, 191. \citet{lnr+03}, 192. \citet{hcm+03b}, 193. \citet{bbj+06}, 194. \citet{mgs03}, 195. \citet{mmp+06}, 196. \citet{tmg+03}, 197. \citet{gpe+03}, 198. \citet{ric+05}, 199. \citet{hcm+03c}, 200. \citet{sgn+03}, 201. \citet{hcs+03}, 202. \citet{lkk+03}, 203. \citet{hsm+03b}, 204. \citet{rsp+04}, 205. \citet{jhf+04}, 206. \citet{bdf+04}, 207. \citet{rsg05}, 208. \citet{rgk+04}, 209. \citet{bss+05}, 210. \citet{b03}, 211. \citet{pmg+05}, 212. \citet{fh03}, 213. \citet{iak+03}, 214. \citet{gdc+03}, 215. \citet{bra+03}, 216. \citet{ctl+03}, 217. \citet{hcv+03}, 218. \citet{sac+03}, 219. \citet{ggm+06}, 220. \citet{pbc+03}, 221. \citet{cbv+04}, 222. \citet{mga+06}, 223. \citet{fsb03}, 224. \citet{mmg+05}, 225. \citet{mpr+04}, 226. \citet{fcd+06}, 227. \citet{mga+05}, 228. \citet{dsg+04}, 229. \citet{fdc+05}, 230. \citet{dra+04}, 231. \citet{gjm+04}, 232. \citet{fdt+04}, 233. \citet{bra+04b}, 234. \citet{dtg+04}, 235. http://space.mit.edu/HETE/Bursts/GRB040810/, 236. \citet{php+04}, 237. \citet{ura+04}, 238. \citet{gdm+04}, 239. \citet{bra+04}, 240. \citet{sbb+07c}, 241. \citet{brv+04}, 242. \citet{aks+07}, 243. \citet{kkt+04}, 244. \citet{fra+04}, 245. \citet{wsr+04}, 246. \citet{huf+05b}, 247. \citet{gra+04}, 248. \citet{bvm+04}, 249. \citet{sgn+05}, 250. http://space.mit.edu/HETE/Bursts/GRB041016/, 251. \citet{kyk04}, 252. \citet{dra+04b}, 253. \citet{cfg04}, 254. \citet{sbb+08}, 255. \citet{bnf+04}, 256. \citet{mhm+06}, 257. \citet{hlb+04}, 258. \citet{nbf+04}, 259. \citet{ebp+07}, 260. \citet{bfk+05}, 261. \citet{rms04}, 262. \citet{r04}, 263. \citet{r04b}, 264. \citet{hms+06}, 265. \citet{dsj+05}, 266. \citet{hra+05}, 267. \citet{lfc+05}, 268. \citet{g+08}, 269. \citet{bck05}, 270. \citet{mts+05}, 271. \citet{lra+05}, 272. \citet{cf05}, 273. \citet{sry05}, 274. \citet{lot+06}, 275. \citet{bg05}, 276. \citet{dfc+05}, 277. \citet{pdl+06}, 278. \citet{nbh+05}, 279. \citet{dfc+05b}, 280. \citet{kb05}. 281. \citet{bm05}, 282. \citet{srm+05}, 283. \citet{jfl+06}, 284. \citet{huk+07}, 285. \citet{tjd+05}, 286. \citet{dbb+06}, 287. \citet{bog05}, 288. \citet{sra+05}, 289. \citet{bgo05}, 290. \citet{dfj+07}, 291. \citet{cf05b}, 292. \citet{kka+05}, 293. \citet{ckg+05}, 294. \citet{hpg+07}, 295. \citet{bmm+05}, 296. \citet{mp05}, 297. \citet{cf05c}, 298. \citet{mbr+05}, 299. \citet{gm05}, 300. \citet{hpo+05}, 301. \citet{pef+05}, 302. \citet{yar+06}, 303. \citet{mgm+05}, 304. \citet{cf05d}, 305. \citet{bpc+06b}, 306. \citet{hop+06}, 307. \citet{fjh+05}, 308. \citet{pra+05}, 309. \citet{cpb+05}, 310. \citet{gjg+05b}, 311. \citet{gmm+05}, 312. \citet{gsm+05}, 313. \citet{gmm+05b}, 314. \citet{chi+05}, 315. \citet{fcb+05b}, 316. \citet{bbb+06}, 317. \citet{mgg+05}, 318. \citet{bb05b}, 319. \citet{gbc+06}, 320. \citet{hkm05}, 321. \citet{gmm+05c}, 322. \citet{mml+05}, 323. \citet{gct+05}, 324. \citet{dbb+05}, 325. \citet{dmg+06}, 326. \citet{mgg+05b}, 327. \citet{bgs+05b}, 328. \citet{gms+05b}, 329. \citet{ksf+05}, 330. \citet{cdm+05}, 331. \citet{rop+07}, 332. \citet{mkh+05}, 333. \citet{atr+06}, 334. \citet{hr05}, 335. \citet{cpb+05b}, 336. \citet{pcm+06}, 337. \citet{dmo+07}, 338. \citet{rmy+06}, 339. \citet{fsj+05}, 340. \citet{odp+07}, 341. \citet{hr05b}, 342. \citet{mra+05}, 343. \citet{rry+05}, 344. \citet{jlf+06}, 345. \citet{jfh+05}, 346. \citet{bgs+05c}, 347. \citet{pbw+05}, 348. \citet{ckh+06}, 349. \citet{pzb+05}, 350. \citet{fjs+05}, 351. \citet{sfg+07}, 352. \citet{mho07}, 353. \citet{h05}. 354. \citet{kka+06}, 355. \citet{hnr+06}, 356. \citet{fcb+05}, 357. \citet{sai+05}, 358. \citet{b05b}, 359. \citet{mcp+06}, 360. \citet{bmh+05}, 361. \citet{gjg+05}, 362. \citet{dp05}, 363. \citet{tjc+05}, 364. \citet{rbp+05}, 365. \citet{rbp+05b}, 366. \citet{bpb+05}, 367. \citet{brg05}, 368. \citet{sbo05}, 369. \citet{sip+05}, 370. \citet{ora+05}, 371. \citet{bdp+05}, 372. \citet{brm+05}, 373. \citet{nys+06}, 374. \citet{rvw+07}, 375. \citet{hra+05b}, 376. \citet{cjp+06}, 377. \citet{mks+05}, 378. \citet{qfh+05}, 379. \citet{pfb+06}, 380. \citet{tbb+05}, 381. \citet{pbf+06}, 382. \citet{ggk+07}, 383. \citet{gpg+07}, 384. \citet{tfj+05}, 385. \citet{mgm+05b}, 386. \citet{pbr+06}, 387. \citet{h06}, 388. \citet{rkb+06}, 389. \citet{aht+06}, 390. \citet{zt06}, 391. \citet{omp+06}, 392. \citet{bdr+06}, 393. \citet{ssa+06}, 394. \citet{kua+06}, 395. \citet{ytk+06}, 396. \citet{mgs+06b}, 397. \citet{ptb06}, 398. \citet{ggm+06b}, 399. \citet{pdf+06}, 400. \citet{ytk+06b}, 401. \citet{kbm06}, 402. \citet{jpk+06}, 403. \citet{cma+06}, 404. \citet{pkc+06}, 405. \citet{pft+06}, 406. \citet{mbs+07}, 407. \citet{gm06}, 408. \citet{njf+06}, 409. \citet{but07}, 410. \citet{cms06}, 411. \citet{cos+06}, 412. \citet{mg06}, 413. \citet{k06}, 414. \citet{kak+06}, 415. \citet{fgj+06}, 416. \citet{hsd+07}, 417. \citet{sdp+07}, 418. \citet{cvb+07}, 419. \citet{cfb06}, 420. \citet{sip+06}, 421. \citet{sip06}, 422. \citet{gam+06b}, 423. \citet{gjc+06}, 424. \citet{mha+06}, 425. \citet{rtm06}, 426. \citet{bkr+06}, 427. \citet{bc06}, 428. \citet{ap06}, 429. \citet{lnh+06}, 430. \citet{hgm06}, 431. \citet{dic06}, 432. \citet{jgg+06}, 433. \citet{cma+06b}, 434. \citet{spk+06}, 435. \citet{bpk06}, 436. \citet{vj06}, 437. \citet{clh+06}, 438. \citet{mht+06}, 439. \citet{tnj+06}, 440. \citet{lgj+06}, 441. \citet{c06b}, 442. \citet{hm06}, 443. \citet{fc06}, 444. \citet{cpf+06}, 445. \citet{hg06}, 446. \citet{ocg+07}, 447. \citet{fwt+06}, 448. \citet{tjf+06}, 449. \citet{h06b}, 450. \citet{p06}, 451. \citet{mgm+06}, 452. \citet{bfk+06}, 453. \citet{m06}, 454. \citet{kba06}, 455. \citet{cbd+06}, 456. \citet{dpm+06}, 457. \citet{dhm+07}, 458. \citet{jhf+06}, 459. \citet{can+06}, 460. \citet{trh06}, 461. \citet{fkk+06}, 462. \citet{sap06}, 463. \citet{lvs+06}, 464. \citet{nrc+07}, 465. \citet{fmj+06}, 466. \citet{vcp+06}, 467. \citet{ssk+06}, 468. \citet{sm06}, 469. \citet{co06}, 470. \citet{jvf+06}, 471. \citet{jfh+06}, 472. \citet{fmc06}, 473. \citet{pmb06}, 474. \citet{mcp06}, 475. \citet{tlj+06}, 476. \citet{pdp+06}, 477. \citet{rp06}, 478. \citet{bbb+06b}, 479. \citet{hcm+06}, 480. \citet{s06}, 481. \citet{cbm06}, 482. \citet{tpb07}, 483. \citet{ltr+06}, 484. \citet{dca06},
485. \citet{mg06b}, 486. \citet{wrt07}, 487. \citet{cr06}, 488. \citet{fdm+06}, 489. \citet{smo06}, 490. \citet{cof06}, 491. \citet{rjt+06}, 492. \citet{wtr06}, 493. \citet{ljh+07}, 494. \citet{hip06}, 495. \citet{mgs+06}, 496. \citet{fro06}, 497. \citet{hc06}, 4989. \citet{cmt06}, 499. \citet{dpc+06}, 500. \citet{ssd+06}, 501. \citet{fjj+06}, 502. \citet{arp06}, 503. \citet{gam+06}, 504. \citet{pch+06}, 505. \citet{tgb+06}, 506. \citet{gmp+06}, 507. \citet{gbm+06}, 508. \citet{cbc+06}, 509. \citet{jfj+06}, 510. \citet{ocp06}, 511. \citet{mmg+07}, 512. \citet{ugh06}, 513. \citet{hm06b}, 514. \citet{mpg+06}, 515. \citet{b06c}, 516. \citet{sbb+06}, 517. \citet{pbb+06}, 518. \citet{kba06b}, 519. \citet{cok06}, 520. \citet{hts+06}, 521. \citet{tfj+06}, 522. \citet{f06}, 523. \citet{fmt+06}, 524. \citet{tmf+06}, 525. \citet{bpc06}, 526. \citet{hma06}, 527. \cite{mpg+06b}, 528. \citet{ha06}, 529. \citet{pbb+08}, 530. \citet{psv+06}, 531. \citet{sms+06}, 532. \citet{kba06c}, 533. \citet{cf06}, 534. \citet{b06b}, 535. \citet{c06c}, 536. \citet{cdy+07}, 537. \citet{bm07}, 538. \citet{jmf+07}, 539. \citet{mjv07}, 540. \citet{fbp+07}, 541. \citet{cf07}, 542. \citet{bhp+07}, 543. \citet{vci07}, 544. \citet{mjv+07}, 545. \citet{cfc+07c}, 546. \citet{hm07}, 547. \citet{dmd+07}, 548. \citet{yyk07}, 549. \citet{mmh07}, 550. \citet{rtm+07}, 551. \citet{tka+07}, 552. \citet{kba07}, 553. \citet{jtf+07}, 554. \citet{rlt+07}, 555. \citet{vrg+07}, 556. \citet{pmg+07}, 557. \citet{mpg+07}, 558. \citet{ha07}, 559. \citet{jfa+07}, 560. \citet{c07}, 561. \citet{meg+07}, 562. \citet{gmk+07}, 563. \citet{jmt+07}, 564. \citet{bmg+07}, 565. \citet{pgh+07}, 566. \citet{cgs+07}, 567. \citet{psv+07}, 568. \citet{djt+07}, 569. \citet{gsb+07}, 570. \citet{cpb+07}, 571. \citet{tjf+07}, 572. \citet{pvr+07}, 573. \citet{jfa+07b}, 574. \citet{vch+07}, 575. \citet{fpb07}, 576. \citet{tgb+07}, 577. \citet{hak07}, 578. \citet{srb+07}, 579. \citet{cpb07}, 580. \citet{ssa+07}, 581. \citet{dcg+07}, 582. \citet{tjf+07b}, 583. \citet{ryy07}, 584. \citet{uph+07}, 585. \citet{cfc+07}, 586. \citet{gpe+07}, 587. \citet{ssd+07}, 588. \citet{sbh+07}, 589. \citet{mtf+07}, 590. \citet{hgk+07}, 591. \citet{mbb+07}, 592. \citet{mrt07}, 593. \citet{bbc+07}, 594. \citet{gbc+07}, 595. \citet{zbs+07}, 596. \citet{pbb+07}, 597. \citet{mjf+07}, 598. \citet{gam+07}, 599. \citet{fcd+07}, 600. \citet{cr07}, 601. \citet{cbb+07}, 602. \citet{ptm+07}, 603. \citet{bm07b}, 604. \citet{sbb+07}, 605. \citet{sp07}, 606. \citet{fbb+07}, 607. \citet{sdk+07}, 608. \citet{tpc+07}, 609. \citet{mbb+07b}, 610. \citet{ks07}, 611. \citet{c07b}, 612. \citet{mbb+07c}, 613. \citet{kba07b}, 614. \citet{cbe+07}, 615. \citet{fsg+07}, 616. \citet{gbp+07}, 617. \citet{jgf+07}, 618. \citet{rb07}, 619. \citet{c07c}, 620. \citet{mpg+07b}, 621. \citet{bpc+07b}, 622. \citet{dsp07}, 623. \citet{ses07}, 624. \citet{pcb+07}, 625. \citet{mgr+07}, 626. \citet{rp07}, 627. \citet{grm+07}. 628. \citet{rmp07}, 629. \citet{ppm+07}, 630. \citet{mkm+07}, 631. \citet{spr+07}, 632. \citet{ilu07}, 633. \citet{mbb+07e}, 634. \citet{im07}, 635. \citet{pbc+07}, 636. \citet{khf07}, 637. \citet{kbc+07}, 638. \citet{xzq+07d}, 639. \citet{hsb+07}, 640. \citet{jvh+07}, 641. \citet{xzq+07b}, 642. \citet{spp+07}, 643. \citet{tpa+07}, 644. \citet{prk+07}, 645. \citet{mpy+07}, 646. \citet{rcs07}, 647. \citet{xzq+07c}, 648. \citet{gcs+07}, 649. \citet{sbs+07}, 650. \citet{ljj+07}, 651. \citet{liu07}, 652. \citet{mgu+07}, 653. \citet{psk+07}, 654. \citet{m07}, 655. \citet{spk+07}, 656. \citet{chu07}, 657. \citet{ukg+06}, 658. \citet{b07c}, 659. \citet{cmd07}, 660. \citet{yys+07}, 661. \citet{seb+07}, 662. \citet{cfo07}
}